\documentclass{amsart}
\usepackage{graphicx,amsmath,amssymb,pxfonts}
\makeatletter
\def\@thmcountersep{.}
\makeatother

\newtheorem{thm}{Theorem}[section]

\newtheorem{lem}[thm]{Lemma}
\newtheorem{rem}[thm]{Remark}
\newtheorem{assu}[thm]{Condition}

\newtheorem{ex}[thm]{Example}
\begin{document}

\title[Edgeworth expansion]{
Asymptotic analysis for stochastic volatility:\\
 Edgeworth expansion}
\date{2010-04-11}
\author{M. Fukasawa}
\thanks{
Masaaki Fukasawa \\ 
Center for the Study of Finance and Insurance, Osaka University \\
Japan Science and Technology Agency\\
1-3, Machikaneyama, Toyanaka, Osaka, 560-8531,
Japan\\
This work was partially supported by KAKENHI 21740074 (MEXT), 
Mirai Labo (Osaka Univ.),\\
Cooperative Research Program (ISM) and CREST (JST)}

\date{2010-04-11}

\setlength{\baselineskip}{6mm}
\begin{abstract}
\setlength{\baselineskip}{5mm}
The validity of an approximation  formula for 
European option prices under a general stochastic volatility model
 is proved
in the light  of the  Edgeworth expansion
for ergodic diffusions.
The asymptotic expansion is around the Black-Scholes price
and is uniform in bounded payoff functions.
The result provides a validation of
an existing singular perturbation expansion formula for
the fast mean reverting stochastic volatility model. 
\end{abstract}


\begin{flushleft}

\end{flushleft}
\begin{flushleft}
\end{flushleft}

\maketitle

\section{Introduction}
In the last decade, many results on asymptotic expansions of option
prices for stochastic volatility models appeared in the literature. 
Such an expansion formula gives an approximation to   
theoretical price of option and sheds light to the shape of  
theoretical implied volatility surface.
See e.g., Gatheral~\cite{Gatheral} for a practical guide.
The primary objective of this article is not 
to introduce a new expansion formula
but 
to prove the validity of an existing one
which was introduced by Fouque et al.~\cite{fmr1}.
We suppose that the log price process $Z$ satisfies
the stochastic differential equation
\begin{equation} \label{SV}
\begin{split}
&\mathrm{d}Z_t = \left\{ r_t - \frac{1}{2} \varphi(X_t)^2 \right\}
 \mathrm{d}t
+ \varphi(X_t) \left[\rho(X_t)\mathrm{d}W^1_t + 
\sqrt{1-\rho(X_t)^2}\mathrm{d}W^2_t \right] \\
& \mathrm{d}X_t = b(X_t) \mathrm{d}t + c(X_t)\mathrm{d}W^1_t
\end{split}
\end{equation}
under a risk-neutral probability measure, 
where $(W^1,W^2)$ is a 2-dimensional standard Brownian motion,
$r = \{r_t \}$ stands for interest rate and is assumed to be
deterministic,
and $b$, $c$, $\varphi$, $\rho$ are Borel functions with $|\rho|\leq 1$.
Under mild conditions on the ergodicity of $X$, 
we validate an approximation
\begin{equation} \label{approx}
D\mathbb{E}[f(Z_T)] \approx D\mathbb{E}[(1+p(N))f(Z_0-\log(D)-\Sigma/2  + \sqrt{\Sigma}N)]
\end{equation}
for every bounded Borel function $f$, where $N \sim \mathcal{N}(0,1)$,
$\Sigma = \Pi[\varphi^2]T$ with the ergodic distribution $\Pi$ of $X$
 and
\begin{equation} \label{poly}
\begin{split}
&D = \exp\left\{ -\int_0^Tr_s \mathrm{d}s\right\}. \\
&p(z) = \alpha  \left\{ 1-z^2 + 
\frac{1}{\sqrt{\Sigma}}(z^3-3z)\right\}, \\ 
&\alpha = -\int_{-\infty}^{\infty} 
\int_{-\infty}^x \left\{\frac{\varphi(v)^2}{\Pi[\varphi^2]} 
-1\right\}\Pi(\mathrm{d}v) \frac{\varphi(x)\rho(x)}{c(x)} \mathrm{d}x.
\end{split}
\end{equation}
In particular, we have a simple formula
\begin{equation*}
D\mathbb{E}[(K-\exp(Z_T))_+] \approx 
P_{\mathrm{BS}}(K,\Sigma) - \alpha d_2(K,\Sigma)
DK\phi(d_2(K,\Sigma))
\end{equation*}
for put option price with strike $K$, where 
$P_{\mathrm{BS}}(K,\Sigma)$ is the Black-Scholes price of 
the put option
\begin{equation*}
\begin{split}
&P_{\mathrm{BS}}(K,\Sigma) = DK\Phi(-d_2(K,\Sigma)) 
- \exp(Z_0)\Phi(-d_2(K,\Sigma)-\sqrt{\Sigma}),
\\
&d_2(K,\Sigma) = - \frac{\log(K)-Z_0+\log(D)}{\sqrt{\Sigma}}
-\frac{\sqrt{\Sigma}}{2}.
\end{split}
\end{equation*}
Notice that if $\alpha= 0$ then 
the right hand side of (\ref{approx}) coincides with
the Black-Scholes price for the European payoff function $f \circ \log$
with volatility $\Pi[\varphi^2]^{1/2}$.
The term with $p$ is small if $c$ is large,
so that in such a  case it
 should be regarded as  a correction term to the  
Black-Scholes approximation. 
The right hand side of  (\ref{approx}) is an alternative representation of
 the so-called fast mean reverting or singular
perturbation  expansion formula
and its validity has been discussed by  Fouque et al.~\cite{fmr2}\cite{fmr5},
Conlon and Sullivan~\cite{CS2005} and 
Khasminskii and Yin~\cite{KY2005} under restrictive conditions on 
the payoff function $f$ or on the coefficients 
of the stochastic differential equation (\ref{SV}).
Recently, Fukasawa~\cite{fands} gave a general framework based on
Yoshida's theory of martingale expansion to 
prove the validity of such an asymptotic expansion around 
the Black-Scholes price for a general 
stochastic volatility model with jumps, 
which in particular 
incorporates the fast mean reverting case with (\ref{SV}).  
This paper, on the other hand, 
concentrates on the particular standard model to improve the preceding
results mainly in the following points:
\begin{enumerate}
\item conditions on the integrability of $\langle Z \rangle$ are weakened,
\item precise order estimate of approximation error is given. 
\end{enumerate}
The framework of Fukasawa~\cite{fands} is too general
to give such a precise  estimate of order of error.
A PDE approach taken by Fouque et al.~\cite{fmr2}\cite{fmr5} gave
order estimates which depend on the regularity of the payoff $f$.
The order given in this article is more precise  and does not depend on 
the regularity of $f$. We require no condition on the smoothness of $f$
and a weaker condition on 
the coefficients $\varphi$, $\rho$, $b$ and $c$.
 We exploit Edgeworth expansion
for ergodic diffusions developed by Fukasawa~\cite{F2008}.

The Edgeworth expansion is a refinement of the central limit theorem
and has played an important role in statistics.
There are three approaches to validate the Edgeworth expansion 
for ergodic continuous-time processes.
Global(martingale) and local(mixing) approaches which were
developed by Yoshida~\cite{martingale} and \cite{partialmixing}  
respectively
are widely applicable
to general continuous-time processes.
The third approach, which is called 
regenerative approach and was developed by Fukasawa~\cite{F2008}
extending Malinovskii~\cite{Malinovskii},
is applicable only to strong Markov processes but
requires weaker conditions of ergodicity and integrability.
The martingale approach was applied
to the validation problem of perturbation expansions 
by Fukasawa~\cite{fands} as noted above.
The present article  exploits the regenerative approach 
which enables us  to treat such an ergodic diffusion 
$X$ that is not  geometrically mixing. 
An extension to this direction  is important because empirical studies
such as
Andersen et al.~\cite{ABDL}
showed that the volatility process appears ``very slowly mean reverting'',
that is, the autocorrelation function decays slowly.
Our model (\ref{SV}) with a condition of ergodicity  given below 
is a natural extension of the fast mean reverting
model of Fouque et al.~\cite{fmr1}\cite{fmr2} but does not necessarily imply a fast decay of
the autocorrelation function.
It admits a polynomial decay of  $\alpha$-mixing coefficient.

It should be noted that our approach in this article utilizes the fact
that $X$ is one-dimensional in (\ref{SV}).
See Fukasawa~\cite{fands} for multi-dimensional fast mean reverting
stochastic volatility model with jumps.
In Section~2, we review the fast mean reverting expansion 
technique. The main result is stated in Section~3 with examples.
An introduction to the Edgeworth expansion theory is given in
Section~4 and then, the proof of the main result is presented in
Section~5. The proof of an important lemma is deferred to Section~6.

\section{Fast mean reverting stochastic volatility}

\subsection{PDE approach}
Here we review an asymptotic method introduced by
Fouque et al.~\cite{fmr1}, where
a family of the stochastic volatility models
\begin{equation} \label{papa}
\begin{split}
&\mathrm{d}S_t^{\eta} = rS_t^{\eta}\mathrm{d}t + \varphi(X_t^{\eta})
S_t^{\eta}\mathrm{d}W^{\rho}_t, \\
&\mathrm{d}X_t^{\eta} = \left\{\frac{1}{\eta^2}
(m-X_t^{\eta}) - \frac{\nu \sqrt{2} }{\eta}
\Lambda(X_t^{\eta})
\right\} \mathrm{d}t + \frac{\nu \sqrt{2}}{\eta}\mathrm{d}W_t
\end{split}
\end{equation}
is considered,
where $W = (W_t)$ and $W^{\rho} = (W^{\rho}_t)$ are standard Brownian motions 
with correlation $\langle W, W^{\rho} \rangle_t = \rho t$,
$\rho \in [-1,1]$.
This is a special case of (\ref{SV}) with
$\rho(x) \equiv \rho$, $b(x) = (m-x)/\eta^2 -
\nu\sqrt{2}\Lambda(x)/\eta$, 
$c(x) \equiv \nu \sqrt{2}/\eta$,
where $m$, $\nu$ are constants
and $\Lambda$ is a Borel function associated with the market price of volatility risk.
For a given payoff function $f$ and maturity $T$, 
the European option price at time $t < T$ defined as
\begin{equation} \label{price}
P^{\eta}(t,s,v) = e^{-r(T-t)}\mathbb{E}[f(S_T^{\eta})|S_t^{\eta} = s,
X_t^{\eta } =  v]
\end{equation}
satisfies 
\begin{equation*}
\left(\frac{1}{\eta^2}\mathcal{L}_0 + 
\frac{1}{\eta}\mathcal{L}_1 + 
\mathcal{L}_2
 \right) P^{\eta}  = 0,
\ \ P^{\eta}(T,s,v) = f(s) 
\end{equation*}
where
\begin{equation*}
\begin{split}
&\mathcal{L}_0 = \nu^2 \frac{\partial^2}{\partial v^2} + 
(m-v)\frac{\partial}{\partial v},\\
& \mathcal{L}_1=\sqrt{2} \rho \nu s \varphi(v)\frac{\partial^2}{
\partial s \partial v} - \sqrt{2}\nu \Lambda(v)\frac{\partial}{\partial
 v}, \\
&\mathcal{L}_2 = \frac{\partial }{\partial t} + 
\frac{1}{2}\varphi(v)^2s^2 \frac{\partial^2}{\partial s^2} + 
r(s\frac{\partial }{\partial s} - 1).
\end{split}
\end{equation*}
Notice that $\mathcal{L}_0$ is the infinitesimal generator
of the OU process 
\begin{equation} \label{ou}
\mathrm{d}X^0_t = (m-X^0_t)\mathrm{d}t + \nu\sqrt{2} \mathrm{d}W_t
\end{equation}
and $\mathcal{L}_2$ is 
the Black-Scholes operator with volatility level $|\varphi(v)|$.
By formally expanding $P^\eta$ in terms of $\eta$ and
equating the same order terms of $\eta$ in the PDE, one
obtains
\begin{equation}\label{sing}
P^{\eta} = P_0 + \eta P_1 + \text{ higher order terms of } \eta
\end{equation}
for the Black-Scholes price $P_0$ with constant volatility
$\Pi_0[\varphi^2]^{1/2}$, where $\Pi_0$ is the ergodic distribution of the 
OU process $X^0$, and 
\begin{equation} \label{f1}
P_0 + \eta P_1
= P_0 - (T-t)\left(
V_2 s^2 \frac{\partial^2 P_0}{\partial s^2} + V_3 s^3
\frac{\partial^3P_0}{\partial s^3}
\right)
\end{equation}
with constants $V_2$ and $V_3$ which are of $O(\eta)$.

As a practical application, 
Fouque et al.~\cite{fmr1} proposed 
its use in calibration problem. They
derived an expansion 
of the Black-Scholes implied volatility $\sigma_{\mathrm{BS}}$ of the form
\begin{equation} \label{iv}
\sigma_\mathrm{BS}(K,T-t) \approx a \frac{\log(K/S)}{T-t} + b
\end{equation}
from (\ref{sing}), 
where $K$ is the  strike price, $S$ is the spot price, $T-t$
is the time to the maturity, 
$a$ and $b$ are constants connecting to $V_2$ and $V_3$ as
\begin{equation} \label{vs}
V_2 = \bar{\sigma}((\bar{\sigma} -b) -a(r +\frac{3}{2}\bar{\sigma}^2)),
\ \ 
V_3 = -a \bar{\sigma}^3, \ \ \bar{\sigma}^2 = \Pi_0[\varphi^2].
\end{equation}
The calibration methodology consists of
(i) estimation of $\bar{\sigma}$ from historical stock  returns,
(ii) estimation of $a$ and $b$ by fitting (\ref{iv}) to
the implied volatility surface, and 
(iii) pricing or hedging by using estimated $\bar{\sigma}$, $a$ and  $b$
via (\ref{f1}) and (\ref{vs}).
This approach captures the volatility skew as well as the term
structure.  It 
enables us to calibrate fast and stably due to parsimony of parameters;
we have no more need to specify all the parameters in the underlying
stochastic volatility model. 
The first step (i) can be eliminated because 
the number of essential parameters is 2 in light of (\ref{approx});
by using $\Pi_{\eta}[\varphi^2]^{1/2}$ instead of
$\Pi_0[\varphi^2]^{1/2}$ for $\bar{\sigma}$, 
where $\Pi_{\eta}$ is the
ergodic distribution of $X^{\eta}$, we can see that 
the right hand side of
(\ref{f1}) coincides with that of  (\ref{approx}) with
$V_3 = -\alpha\Pi_{\eta}[\varphi^2]$ and $V_2 = 2V_3$.

It should be explained 
what is the intuition of $\eta \to 0$.
To fix ideas, let $\Lambda = 0$ for brevity.
Then $\tilde{X}_t := X_{\eta^2 t}^{\eta}$ satisfies
\begin{equation*}
\mathrm{d}\tilde{X}_t = 
(m-\tilde{X}_t)\mathrm{d}t + \nu \sqrt{2}\mathrm{d}\tilde{W}_t,
\end{equation*}
where $\tilde{W}_t = \eta^{-1}W_{\eta^2 t}$ is a standard 
Brownian motion, and it holds 
\begin{equation*}
\mathrm{d}S_t^{\eta} = r S_t^{\eta} \mathrm{d}t + 
\varphi(\tilde{X}_{t/\eta^2}) 
S_t^{\eta} \mathrm{d}W^{\rho}_t.
\end{equation*}
Hence $\eta$ stands for the volatility time scale.
Note that
\begin{equation*}
\langle \log(S^{\eta}) \rangle_t =  
\int_0^t f(\tilde{X}_{s/\eta^2})^2\mathrm{d}s 
\sim \eta^2 \int_0^{t/\eta^2} \varphi(X^0_s)^2\mathrm{d}s 
\to \Pi_0[\varphi^2] t
\end{equation*}
by the law of large numbers for ergodic diffusions,
where $X^0$ is a solution of (\ref{ou}).
This convergence implies that the log price $\log(S_t^\eta)$
is asymptotically normally distributed with mean 
$rt -\Pi_0[\varphi^2]t/2$ and 
variance $\Pi_0[\varphi^2]t$ by martingale central limit theorem.
The limit is nothing but the Black-Scholes model
with volatility $\Pi_0[\varphi^2]^{1/2}$.
The asymptotic expansion formula around the Black-Scholes price 
can be therefore regarded as a refinement of
a normal approximation based on the central limit theorem
for ergodic diffusions.

\subsection{Martingale expansion}
Note that a formal calculation as in (\ref{sing})
does not ensure in general 
that the asymptotic expansion formula is actually valid. 
A rigorous validation is not easy if the payoff $f$
or a coefficient of the stochastic differential equation is not smooth.
See e.g. Fouque et al.~\cite{fmr2}.
A general result on the validity is given by Fukasawa~\cite{fands}.
Here we state a simplified version of it.
Consider a sequence of models of type (\ref{SV}):
\begin{equation*}
\begin{split}
&\mathrm{d}Z^n_t = \left\{ r_t - \frac{1}{2} \varphi(X^n_t)^2 \right\}
 \mathrm{d}t
+ \varphi(X^n_t) \left[\rho(X^n_t)\mathrm{d}W^1_t + 
\sqrt{1-\rho(X^n_t)^2}\mathrm{d}W^2_t \right] \\
& \mathrm{d}X^n_t = b_n(X^n_t) \mathrm{d}t + c_n(X^n_t)\mathrm{d}W^1_t,
\end{split}
\end{equation*}
where $b_n$ and $c_n$, $n \in \mathbb{N}$ are sequences of Borel
functions. 
\begin{thm}
Suppose that for any $p > 0$, the $L^p$ moments of
\begin{equation} \label{unifint}
\int_0^T \varphi(X^n_t)^2\mathrm{d}t, \ \ 
\left\{ \int_0^T \varphi(X^n_t)^2(1-\rho(X^n_t)^2)\mathrm{d}t
\right\}^{-1}
\end{equation}
are bounded in $n \in \mathbb{N}$ and that
there exist  positive sequences $\epsilon_n$,  $\Sigma_n$ with
$\epsilon_n \to 0$,  $\Sigma_\infty := \lim_{n\to\infty}\Sigma_n > 0$
such that
\begin{equation} \label{convN}
\left(\frac{M^n_T}{\sqrt{\Sigma_n}}, \frac{\langle M^n \rangle_T -
 \Sigma_n}{\epsilon_n \Sigma_n}\right) \to \mathcal{N}(0,V)
\end{equation}
in law with a $2\times 2$ variance 
matrix $V = \{V_{ij}\}$ as $n \to \infty$, where
$M^n$ is the local martingale part of $Z$. Then, for every Borel
function $f$ of polynomial growth,
\begin{equation}\label{martexp}
\mathbb{E}[f(Z^n_T)] = \mathbb{E}[(1+p_n(N))f(Z_0-\log(D)-\Sigma_n/2
 +\sqrt{\Sigma_n}N)]
+ o(\epsilon_n)
\end{equation}
as $n \to \infty$, where $N \sim \mathcal{N}(0,1)$, $D$ is
defined as in (\ref{poly}) and 
\begin{equation*} 
p_n(z) = \epsilon_n \frac{V_{12} }{2}  \left\{ -\sqrt{\Sigma_n}(z^2-1) + 
(z^3-3z)\right\}.
\end{equation*}
\end{thm}

An appealing point of this theorem is that it gives a validation of
not only the singular perturbation but also regular perturbation
expansions including the so-called small vol-of-vol expansion.
It is also noteworthy that the asymptotic skewness $V_{12}$
appeared in the expansion formula is represented as
the asymptotic covariance between the log price and 
the integrated volatility.
Our interest here is however to deal with the singular case only. 
Now, suppose that
$b_n$ and $(1+c_n^2)/c_n$ are locally integrable and locally
bounded on $\mathbb{R}$ respectively for each $n \in \mathbb{N}$;
we take $\mathbb{R}$ as  the state space of $X^n$
by a suitable scale transformation.
Further, we assume that
$s_n(\mathbb{R}) = \mathbb{R}$ for each $n \in \mathbb{N}$, which
ensures that there exists a unique weak solution of (\ref{SV}).
See e.g., Skorokhod~\cite{Sko}, Section~3.1.
It is also known that 
the ergodic distribution $\Pi_n$ of $X^n$ is, if exists, 
given by
\begin{equation*}
\Pi_n(\mathrm{d}x) = \frac{\mathrm{d}x}{ \epsilon_n^2 s_n^\prime(x) c_n^2(x)}, \  \
s_n(x) = 
\int_{0}^x  \exp \left\{-2 \int_{0}^v \frac{b_n(w)}{c_n(w)^2} 
\mathrm{d}w\right\} \mathrm{d}v
\end{equation*}
with a normalizing constant $\epsilon_n^2$:
\begin{equation*}
\epsilon^2_n  = \int \frac{\mathrm{d}x}{ s_n^\prime(x) c_n^2(x)}.
\end{equation*}

\begin{thm}
Suppose that
\begin{enumerate}
\item for any $p > 0$, the $L^p$ boundedness of the sequences
      (\ref{unifint}) holds,
\item $\epsilon_n \to 0$ as $n\to \infty$,
\item $\lim_{n\to \infty\ }\Pi_n[\varphi^2]$  exists and is positive,
\item $\lim_{n\to \infty\ }\Pi_n[\varphi \rho \psi_n  ]$ and  
$\lim_{n\to \infty\ }\Pi_n[\psi_n^2 ]$ exist, where
\begin{equation*}
\psi_n(x) = 2\epsilon_n c_n(x)s_n^{\prime}(x)\int_{-\infty}^{x}
(\varphi(\eta)^2 - \Pi_n[\varphi^2])\Pi_n(\mathrm{d}\eta),
\end{equation*}
\item  
the sequences
\begin{equation*}
\int_{X^n_0}^{X^n_T}\frac{\psi_n(x)}{c_n(x)} \mathrm{d}x, \ \ 
\frac{1}{T}\int_0^T \psi_n(X^n_t)^2 \mathrm{d}t - \Pi_n[\psi_n^2]
\end{equation*}
and
\begin{equation*} 
\frac{1}{T}\int_0^T \psi_n(X^n_t)\rho(X^n_t)\varphi(X^n_t)\mathrm{d}t - 
\Pi_n[\psi_n \rho \varphi]
\end{equation*}
converge to $0$ in probability as $n\to \infty$.
\end{enumerate}
Then, the approximation (\ref{approx}) is valid in that
(\ref{martexp}) holds with $\Sigma_n  = \Pi_n[\varphi^2]T$ and
 $p_n = p$ defined as
 (\ref{poly})
with $b=b_n$, $c=c_n$ and $\Sigma = \Sigma_n$.
\end{thm}
{\it Proof: }
Let us verify
(\ref{convN}) with $\Sigma_n  = \Pi_n[\varphi^2]T$ and
\begin{equation*}
V_{12} = -2 \lim_{n\to \infty }\Sigma_n^{-3/2} \Pi_n[\varphi \rho \psi_n].
\end{equation*}
Notice that by the It$\hat{\text{o}}$-Tanaka formula,
\begin{equation*}
\begin{split}
\langle M^n \rangle_T - \Pi_n[\varphi^2]T =& 
\int_0^T (\varphi(X^n_t)^2 - \Pi_n[\varphi^2]) \mathrm{d}t\\  
&= \epsilon_n \int_{X^n_0}^{X^n_T} \frac{\psi_n(x)}{c_n(x)} \mathrm{d}x -
\epsilon_n \int_0^T \psi_n(X^n_t)\mathrm{d}W^1_t.
\end{split}
\end{equation*}
It suffices then to prove the asymptotic normality of
\begin{equation*}
\left(
\int_0^T \varphi(X^n_t)\left[\rho(X^n_t)\mathrm{d}W^1_t + \sqrt{1-\rho(X^n_t)^2}
\mathrm{d}W^2_t\right], 
\int_0^T \psi_n(X^n_t)\mathrm{d}W^1_t
\right).
\end{equation*}
This follows from the martingale central limit theorem under
the fifth assumption.
\hfill////

The conditions are easily verified in such a case
that both $\Pi_n$ and $s_n$ do not depend on $n \in \mathbb{N}$.
The model (\ref{papa}) with $\Lambda\equiv 0$
and $\eta = \eta_n$, where $\eta_n$ is a positive sequence with
$\eta_n \to 0$,
is an example of such an easy case.

\section{Main results}
\subsection{Main theorem and remarks}
Here we state the main results of this article. 
We treat (\ref{SV}) with 
Borel functions $\varphi$, $\rho$ satisfying
$|\rho|\leq 1$, $b$ being a locally integrable function on
$\mathbb{R}$ and $c$ being a positive Borel function such that
$(1+c^2)/c$ is locally bounded on $\mathbb{R}$.
We suppose that $\varphi$ also is locally bounded on $\mathbb{R}$
and that there exists a non-empty open set $U \subset \mathbb{R}$ such
that
it holds on $U$ that
\begin{enumerate}
\item  $\varphi$ and  $\rho$ are continuously differentiable,
\item $(1-\rho^2)\varphi^2 > 0$ and  $|\varphi^\prime| > 0$.
\end{enumerate}
If $\varphi$ is constant, then the approximation (\ref{approx})
is trivially valid.
Since $U$ can be any open set as long as it is not empty, 
this condition is not restrictive in the context of stochastic
volatility models.
This rules out, however, the case $\rho \equiv 1$.
We can introduce alternative framework to include such a
case although 
 we do not go to the details in this article for the sake of brevity.
We fix $\varphi$, $\rho$, $U$ and  assume
$(Z_0, X_0) = (0,0)$ without loss of generality.

Define the scale function $s : \mathbb{R} \to \mathbb{R}$
and the normalized speed measure density
  $\pi: \mathbb{R} \to \mathbb{R}$ 
as
\begin{equation} \label{pi}
s(x) = 
\int_{0}^x  \exp \left\{-2 \int_{0}^v \frac{b(w)}{c(w)^2} 
\mathrm{d}w\right\} \mathrm{d}v, \ \ 
\pi(x) = \frac{1}{ \epsilon^2 s^\prime(x) c^2(x)}
\end{equation}
with
\begin{equation} \label{ep}
\epsilon^2  = \int \frac{\mathrm{d}x}{ s^\prime(x) c^2(x)}.
\end{equation}
It is well-known that the stochastic differential equation for $X$ in
(\ref{SV}) has a unique weak solution which 
is ergodic if $\epsilon < \infty$ and
$s(\mathbb{R}) = \mathbb{R}$.
The ergodic distribution $\Pi$ of $X$ is given by
$\Pi(\mathrm{d}x) = \pi(x)\mathrm{d}x$.
See e.g., Skorokhod~\cite{Sko}, Section~3.1.
Notice that $X$ is completely characterized by 
$(\pi, s, \epsilon)$.
In fact, we can recover $b$ and $c$ by $1/c^2 = \epsilon^2 s^\prime \pi$
and $b = -c^2 s^{\prime \prime}/2s^\prime$. Taking this into mind,
denote by $\mathcal{C}$ the set of
all triplets $(\pi,s,\epsilon)$ with $\pi$ being
a locally bounded probability density function on $\mathbb{R}$ such that
$1/\pi$ is also locally bounded on $\mathbb{R}$,
$s$ being a bijection from $\mathbb{R}$ to $\mathbb{R}$
such that $s^\prime$ exists and is a positive absolutely continuous function,
and $\epsilon$ being a positive finite constant.

For given $\gamma = (\gamma_+, \gamma_-) \in [0,\infty)^2$ 
and $\delta  \in (0,1)$, 
denote by $\mathcal{C}(\gamma, \delta)$
the set of $\theta = (\pi, s, \epsilon) \in \mathcal{C}$ satisfying 
Conditions~\ref{assgam}, \ref{assint} below.

\begin{assu} \label{assgam}
It holds that
\begin{equation*}
(1+\varphi(x)^2)\pi(x)s^{\prime}(y) \leq \exp\{
-\log(\delta) + \gamma_+x - (4\gamma_++\delta)(x-y)\}
\end{equation*}
for all $x \geq y \geq 0$ and
\begin{equation*}
(1+\varphi(x)^2)\pi(x)s^{\prime}(y) \leq \exp\{
-\log(\delta) - \gamma_-x + (4\gamma_-+\delta)(x-y)\}
\end{equation*}
for all $x \leq y \leq 0$.
\end{assu}

\begin{assu}\label{assint}
There exist $x \in U$ and  $a \in [\delta,1/\delta]$ such that
$|x|\leq 1/\delta$, 
$[x-a,x+a] \subset U$, 
$\pi$ is absolutely continuous on $[x-a,x+a]$ and
it holds
\begin{equation*}
\sup_{y \in [x-a,x+a]}
\left|\left(\sqrt{\frac{\pi}{s^\prime}}\varphi \rho \right)^\prime(y)\right|
\vee s^\prime(y) 
\vee \pi(y) \vee \frac{1}{s^\prime(y)} \vee
\frac{1}{\pi(y)} 
\leq 1/\delta.
\end{equation*}
\end{assu}

Given $\theta \in \mathcal{C}$, 
we write $\pi_{\theta}, s_{\theta}, \epsilon_\theta, 
b_\theta, c_{\theta}, Z^\theta$ 
for the elements of $\theta = (\pi, s, \epsilon)$,
the corresponding coefficients $b$, $c$ 
of the stochastic differential equations,
and the log price process $Z$ defined as (\ref{SV}) respectively.

\begin{thm}\label{main}
Fix $\gamma  = (\gamma_+, \gamma_-) \in [0,\infty)^2$
and $\delta \in (0,1)$. Denote by $\mathcal{B}_\delta$ 
the set of the Borel functions bounded by $1/\delta$.
Then,
\begin{equation*}
\sup_{f \in \mathcal{B}_\delta, \theta \in \mathcal{C}(\gamma,\delta)}
\epsilon_{\theta}^{-2}\left|
\mathbb{E}[f(Z^\theta_T)]- \mathbb{E}[(1+p_\theta(N))
f(-\log(D)-\Sigma_\theta/2  + \sqrt{\Sigma_\theta}N)]
\right|
\end{equation*}
is finite, where $N\sim \mathcal{N}(0,1)$, 
$\Sigma_\theta = \Pi_\theta[\varphi^2]T$, 
$\Pi_\theta(\mathrm{d}x) = \pi_\theta(x)\mathrm{d}x$ 
and $D$, $p=p_\theta$ are defined by (\ref{poly}) with 
$\Sigma =\Sigma_\theta$, $\Pi = \Pi_\theta$, $c = c_\theta$.
\end{thm}

\begin{rem} \label{etarem} \upshape
The point of the definition of $\mathcal{C}(\gamma,\delta)$
is that it is written independently of $\epsilon$. As a result,
if $\theta \in \mathcal{C}(\gamma, \delta)$, then
$(\pi_\eta, s_\eta,\epsilon_\eta)$ associated with the drift coefficient
$b_\eta= b_\theta/\eta^2$ and the diffusion coefficient 
$c_\eta = c_\theta/\eta$
is also an element of $\mathcal{C}(\gamma,\delta)$ 
for any $\eta > 0$. In fact $\pi_\eta = \pi_\theta$ and 
$s_{\eta} = s_\theta$.
On the other hand, $\epsilon_{\eta} = \eta \epsilon_{\theta}$, so
that Theorem~\ref{main} implies, with a slight abuse of
notation,
\begin{equation} \label{eta}
E[f(Z^\eta_T)]= E[(1+p_\eta(N))
f(Z_0-\log(D)-\Sigma_\eta/2  + \sqrt{\Sigma_\eta}N)] +O(\eta^2)
\end{equation}
as $\eta \to 0$.
\end{rem}

\begin{rem}\upshape
Given $\theta \in \mathcal{C}$,
Condition~\ref{assint} does not hold for any  $\delta > 0$
only when considering vicious examples such as the case
$(\sqrt{\pi_\theta/s_\theta^\prime}\varphi \rho)^\prime $ is not continuous at
any point of $U$;
a sufficient condition for Condition~\ref{assint} to hold with some
$\delta > 0$ is that
$(\sqrt{\pi_\theta/s_\theta^\prime}\varphi \rho)^\prime $
is continuous at  some point of $U$. 
If Condition~\ref{assint} holds with some $\delta > 0$, then it holds
with any $\hat{\delta} \in (0,\delta]$ as well. 
\end{rem}


\subsection{Examples}

\begin{lem} \label{assklem}
Let $\theta \in \mathcal{C}$.
If there exist $(\gamma_{+}, \gamma_-) \in [0,\infty)^2$  such that
\begin{equation}\label{assk}
\kappa_\pm > 2\gamma_\pm, \ \ 
\limsup_{v \to \pm \infty}
 \frac{1 + \varphi(v)^2}
{e^{\gamma_{\pm}|v|} c_\theta(v)^2}
< \infty
\end{equation}
with
\begin{equation*}
\kappa_+ = - \limsup_{v \to \infty}
\frac{b_\theta(v)}{c_\theta(v)^2}, 
\ \ 
\kappa_- = \liminf_{v \to -\infty}
\frac{b_\theta(v)}{c_\theta(v)^2},
\end{equation*}
then 
there exists $\delta_0>0$ such that for any 
$\delta \in (0,\delta_0\wedge 1)$,
Condition~\ref{assgam} holds for $\theta = (\pi,s,\epsilon)$ with
$\gamma = (\gamma_+,\gamma_-)$ and $\delta$.
\end{lem}
{\it Proof: }
This is shown in a straightforward manner by (\ref{pi}).
\hfill////

\begin{ex} \upshape
Consider 
\begin{equation*}
\begin{split}
&\mathrm{d}Z_t = \left\{r_t - \frac{1}{2}V_t \right\}\mathrm{d}t 
+ \sqrt{V_t}(\rho \mathrm{d}W^1_t + 
\sqrt{1-\rho^2}\mathrm{d}W^2_t) )\\
&\mathrm{d}V_t = \xi \eta^{-2}(\mu - V_t)\mathrm{d}t + 
\eta^{-1}|V_t|^{\nu}\mathrm{d}W^1_t
\end{split}
\end{equation*}
for positive constants $\xi, \mu, \eta>0$, $\rho \in (-1,1)$ 
and $\nu \in [1/2 ,\infty)$.
We assume $\xi \mu > 1/2$ if  $\nu = 1/2$.
Then, the scale function $s^V$ of $V$
satisfies $s^V((0,\infty)) = \mathbb{R}$, so that we can apply 
 It$\hat{\text{o}}$'s formula to $X = \log(V)$ to have
\begin{equation*}
\mathrm{d}X_t = \eta^{-2}(\xi  \mu e^{-X_t} -\xi - 
e^{-2(1-\nu)X_t}/2 )\mathrm{d}t
+ \eta^{-1}e^{-(1-\nu)X_t} \mathrm{d}W^1_t.
\end{equation*}
In this scale, $\varphi(x) = \exp(x/2)$, so that we can take
any open set as $U \subset \mathbb{R}$.
We fix $\xi, \mu, \nu, \rho$ arbitrarily.
In the light of Remark~\ref{etarem}, it suffices to verify 
Conditions~\ref{assgam} and \ref{assint} only when $\eta=1$.
It is trivial that Condition~\ref{assint} holds
with a  sufficiently small $\delta > 0$.
If $\nu = 1/2$, then ($\ref{assk}$) also holds with
\begin{equation*}
\kappa_+ = \infty, \ \ \kappa_- =   \xi \mu - \frac{1}{2},
\ \ \gamma_+ = 2, \ \ \gamma_- = 0. 
\end{equation*}
If $\nu \in  (1/2,1)$, then it holds with
\begin{equation*}
\kappa_{\pm} = \infty, \ \ \gamma_+ = 3-2\nu, \ \ \gamma_- = 0. 
\end{equation*}
If $\nu = 1$, it then holds with
\begin{equation*}
\kappa_{+} = \xi + \frac{1}{2},\  \
\kappa_- = \infty,
 \ \ \gamma_+ = 1, \ \ \gamma_- = 0 
\end{equation*}
provided that $\xi > 3/2$.
Unfortunately, (\ref{assk}) does not hold if $\nu \in (1,11/8]$.
If $\nu > 11/8$, it then holds with
\begin{equation*}
\kappa_{+} = \frac{1}{2},\  \
\kappa_- = \infty
 \ \ \gamma_+ = (3-2\nu)_+, \ \ \gamma_- = 2\nu -2. 
\end{equation*}
Note that the case $\nu = 1/2$ corresponds to the Heston model.
In this case, we have a more explicit expression of 
the asymptotic expansion formula; we have (\ref{eta}) with
\begin{equation*}
p_\eta(z)= \frac{\eta \rho }{2\xi}\left\{
1-z^2+\frac{1}{\Sigma_\eta^{1/2}}(z^3-3z)
\right\}, \ \ \Sigma_{\eta} = \mu T.
\end{equation*}
This is due to the fact that the ergodic distribution of the CIR
process is a gamma distribution.
\end{ex}

\begin{ex}\upshape
Here we treat (\ref{papa}).
In order to prove the validity of the singular expansion 
in the form (\ref{eta}) for $Z^\eta_T = \log(S^\eta_T)$, 
it suffices to show that there exist $\gamma$, $\delta$ and $\eta_0 >0$ 
such that
Conditions~\ref{assgam} and \ref{assint} hold for
$\theta = (\pi,s,\epsilon) \in \mathcal{C}$ associated with
\begin{equation*}
b_\theta(x) = m-x - \eta \nu\sqrt{2}\Lambda(x), \ \ c_\theta(x) = \nu \sqrt{2}
\end{equation*}
 for any $\eta \in (0,\eta_0]$, in the light of Remark~\ref{etarem}.
Here we fix $m \in \mathbb{R}$ and $\nu \in (0,\infty)$.
Suppose that there exists $(\gamma_+, \gamma_-) \in [0,\infty)^2$ such that
\begin{equation*}
\limsup_{x\to \pm \infty} e^{-\gamma_{\pm}|x|}\varphi^2(x) < \infty
\end{equation*}
and that  $\Lambda$ is locally bounded on $\mathbb{R}$ with
\begin{equation*}
\lambda_\infty := \liminf_{|x| \to \infty }\frac{\Lambda(x)}{x} > - \infty.
\end{equation*}
Then we have
\begin{equation*}
-\mathrm{sgn}(v)\frac{b_\theta(v)}{c_\theta(v)^2} \to \infty, 
\end{equation*}
as $|v| \to \infty$ uniformly in $\eta \in (0,\eta_0]$ with, say, 
 $\eta_0 = 1 \wedge |1/(2\nu\lambda_\infty \wedge 0) |$.
Hence, there exists $\delta \in (0,1)$ such that
Condition~\ref{assgam} holds for any $\eta \in (0,\eta_0]$ with 
$\gamma = (\gamma_+, \gamma_-)$  and $\delta$.
 By, if necessary, replacing $(\delta, \eta_0)$ with a smaller one,
Condition~\ref{assint} also is verified 
for any $\eta \in (0,\eta_0]$
under a slight condition on $\varphi$ stated in the beginning of this
 section.
Consequently, by Theorem~\ref{main}, we have (\ref{eta}) for (\ref{papa})
if $|\rho| < 1$ in addition.
The obtained estimate of error $O(\eta^2)$ is a stronger result than 
one obtained by Fouque et al.~\cite{fmr2}\cite{fmr5}.
\end{ex}

\begin{ex} \upshape
Here we treat a diffusion which is not geometrically mixing.
Consider the stochastic differential equation
\begin{equation*}
\mathrm{d}X_t= -\frac{1}{\eta^2}
\left(\frac{1}{2}+\xi\right)\frac{\tanh(Y_t)}{\cosh(Y_t)^2} 
\mathrm{d}t + \frac{1}{\eta}\frac{1}{\cosh(X_t)} \mathrm{d}W_t
\end{equation*}
with $\xi > 1/2$ and $\eta > 0$.
Putting $Y_t = \sinh (X_t)$, we have
\begin{equation*}
\mathrm{d}Y_t = -\frac{1}{\eta^2}
\frac{\xi Y_t}{1+Y_t^2}\mathrm{d}t + \frac{1}{\eta}\mathrm{d}W_t
\end{equation*}
This stochastic differential equation
has a unique weak solution which is ergodic. 
A polynomial lower bound for the $\alpha$ mixing coefficient
is given in Veretennikov~\cite{Ver} which implies in particular
that $X = \sinh^{-1}(Y)$ is not geometrically mixing for any $\xi$.
Now, let us verify Conditions~\ref{assgam} and \ref{assint}
for (\ref{SV}) with
\begin{equation*}
b(x) =
 -\frac{1}{\eta^2}\left(\frac{1}{2}+\xi\right)
\frac{\tanh(x)}{\cosh(x)^2}, \ \ 
c(x) = \frac{1}{\eta}\frac{1}{\cosh(x)}
\end{equation*}
for any $\eta>0$. In the light of Remark~\ref{etarem}, 
it suffices to deal with the case $\eta=1$.
Since
\begin{equation*}
-\lim_{|x| \to \infty} \mathrm{sgn}(x)\frac{b(x)}{c(x)^2} = 
\frac{1}{2} + \xi,
\end{equation*}
we have (\ref{assk}) if there exists $\mu \geq 0$ such that
\begin{equation*}
\sup_{|x| \to \infty} e^{-\mu|x|}\varphi(x)^2 < \infty, \ \ 
\frac{1}{2}+\xi > 4+2\mu.
\end{equation*}
Condition~\ref{assint} also is satisfied with a sufficiently small
 $\delta > 0$ if the condition stated at the beginning of this section
 holds.
\end{ex}

\section{Edgeworth expansion}
\subsection{Gram-Charlier expansion}
Here we give a brief introduction to  the Edgeworth expansion. 
It is in a sense a rearrangement of the Gram-Charlier expansion.
Let $Y$ be a random variable with $\mathbb{E}[Y]=0$ and $\mathbb{E}[Y^2]=1$. 
If it has a density $p_Y$ with an integrability 
condition
\begin{equation} \label{L2}
\int p_Y(z)^2 \phi(z)^{-1} \mathrm{d}z < \infty,
\end{equation}
where $\phi$ is the standard normal density, then  we have
\begin{equation*}
p_Y/\phi = \sum_{j=0}^{\infty} \frac{1}{j!}\mathbb{E}[H_j(Y)]H_j
\end{equation*}
in $L^2(\phi)$ with Hermite polynomials $H_j$
 defined as
the coefficients of the Taylor series
\begin{equation}\label{Hermite}
e^{tx-t^2/2} = \sum_{j=0}^{\infty} H_j(x) \frac{t^j}{j!},
\ \  (t,x) \in \mathbb{R}^2.
\end{equation}
This is an orthonormal series expansion of $p_Z/\phi \in L^2(\phi)$ and
implies that
\begin{equation} \label{GS}
\mathbb{E}[f(Y)] = \sum_{j=0}^{\infty} \frac{1}{j!}\mathbb{E}[H_j(Y)] \int
 f(z)H_j(z)\phi(z) \mathrm{d}z
\end{equation}
for $f \in L^2(\phi)$.
The Edgeworth formula is obtained by rearranging this Gram-Charlier
series.
For example, if $Y = m^{-1/2} \sum_{j=1}^mX_j$ with an iid sequence $X_j$, then
the $j$-th cumulant $\kappa^Y_j$ of $Y$ is of $O(m^{1-j/2})$. This is simply because
\begin{equation*}
\partial^j \log(\psi_Y(u))= m \partial^j \log (\psi_X(m^{-1/2}u)),
\end{equation*}
where $\psi_Y$ and $\psi_X$ are the characteristic functions of
$Y$ and $X_j$ respectively. 
Even if $Y$ is not an iid sum, $\kappa^Y_j = O(m^{1-j/2})$ often remains
true in cases where  $Y$ converges in law to a normal distribution as $m \to \infty$.
Since $\mathbb{E}[H_0(Y)]=1$, 
$\mathbb{E}[H_1(Y)] = \mathbb{E}[H_2(Y)] = 0$
and for $j \geq 3$,
\begin{equation*}
\mathbb{E}[H_j(Y)] = \sum_{k=1}^{[j/3]}  
\sum_{r_1 + \dots + r_k = j, \  r_j \geq 3}
 \frac{\kappa^Y_{r_1} \dots \kappa^Y_{r_k}}{r_1! \dots r_k!} \frac{j!}{k!}
\end{equation*}
by (\ref{Hermite}),
it follows from (\ref{GS}) 
that
\begin{equation*}
\mathbb{E}[f(Y)] = \sum_{j=0}^J m^{-j/2}\int f(z)q_j(z) \phi(z) \mathrm{d}z + o(m^{-J/2})
\end{equation*}
with suitable polynomials $q_j$.
Taking $J=0$, we have the central limit theorem; in this sense, the
Edgeworth expansion is a refinement of the central limit theorem. 
This asymptotic expansion can be validated under weaker conditions
than (\ref{L2});
see Bhattacharya and Rao~\cite{BR} and Hall~\cite{Hall} 
for iid cases.
Here we give one of the validity theorems.
\begin{thm} \label{BR}
Let $X_j^n$ be a triangular array of $d$-dimensional 
independent random variables with mean $0$.
Assume that $X_j^n \sim X_1^n$ for all $j$ and that
\begin{equation*}
\sup_{n\in \mathbb{N}}\mathbb{E}[|X_1^n|^\xi] < \infty
\end{equation*}
for an integer  $\xi \geq 4$,
\begin{equation*}
\sup_{|u| \geq b, n \in \mathbb{N}} |\Psi^n(u)| < 1, 
\ \ 
\sup_{n\in \mathbb{N}}  \int_{\mathbb{R}^d} |\Psi^n(u)|^\eta \mathrm{d}u < \infty, 
\end{equation*}
for all $b > 0$ and for some $\eta \geq 1$ respectively, where
\begin{equation*}
\Psi^n(u) = \mathbb{E}[\exp\{iu\cdot X_1^n\}].
\end{equation*}
Then, there exists $m_0$ such that
$S_m^n = m^{-1/2}\sum_{j=1}^mX_j^n$ has a bounded density $p_m^n$
for all $m \geq m_0$, $n \in \mathbb{N}$. Further, it holds that
\begin{equation*}
\sup_{x \in \mathbb{R}^k, m \geq m_0, n \in \mathbb{N}} m(1+|x|^\xi)
|p_m^n(x) - q_m^n(x)| < \infty,
\end{equation*}
where
\begin{equation*}
q_m^n(x) = \phi(x;0,v^n) - \frac{1}{6\sqrt{m}}\sum_{i,j,k=1}^{d}
\kappa_{ijk}^n \partial_i\partial_j\partial_k \phi(x;0,v^n)
\end{equation*}
with the variance matrix $v^n$  of $X_1^n$ and
the third moments  $\kappa_{ijk}^n$ of $X_1^n$.
\end{thm}
{\it Proof: }
This result is a variant of Theorem~19.2 of Bhattacharya and Rao~\cite{BR}. 
Although the distribution of $X_1^n$ depends on $n$, 
the assertion is proved in a similar manner
with the aid of Theorem~9.10 of Bhattacharya and Rao~\cite{BR}, due to
our  assumptions. For example, we have ( and use )
\begin{equation*}
0 <
\inf_{|u| = 1, n \in \mathbb{N}} \mathbb{E}[|u\cdot X_1^n|^2]
 \leq \sup_{|u| = 1, n \in \mathbb{N}} \mathbb{E}[|u\cdot X_1^n|^2] < \infty.
\end{equation*}
\hfill////

\subsection{Edgeworth expansion for regenerative functionals}
We have seen that the fast mean reverting expansion
gives a correction term to the Black-Scholes price that corresponds to 
the central limit of an additive functional of ergodic diffusion
in Section~2.1.
In order to prove the validity of the expansion,
it is therefore natural to apply the Edgeworth expansion theory
for ergodic diffusions.
Here we present a general result for 
triangular arrays of regenerative functionals, 
which extends a result for additive functionals of 
ergodic diffusions given by
Fukasawa~\cite{F2008}.
Let $\mathbb{P}^n = (\Omega^n, \mathcal{F}^n, \{\mathbb{F}^n_t\}, P^n)$
be a  family of filtered probability spaces
satisfying the usual assumptions and 
$K^n = (K^n_t)$ 
be an $\{\mathbb{F}^n_t\}$-adapted 
cadlag process defined on 
$\mathbb{P}^n$. 
Denote by $E^n[\cdot]$ and $\mathrm{Var}^n[\cdot]$ the expectation and variance
with respect to $P^n$ respectively.
For a given sequence of increasing $\{\mathbb{F}_t^n\}$-stopping times
$\{\tau_j^n\}$ with $\tau^n_0 = 0$ and 
$\lim_{j\to\infty}\tau_j^n = \infty$, 
put
\begin{equation*}
\mathcal{K}^n_j = \left(\mathcal{K}^n_{j,t}\right)_{ 
t \geq 0}, \ \ 
\mathcal{K}^n_{j,t} = K^n_{t + \tau_j^n } -
		       K^n_{\tau_j^n}, \ \ 
l_j^n = \tau_{j+1}^n - \tau_j^n, \ \ 
j=0,1,2, \dots
\end{equation*}
We say that $K^n$ is a {\bf regenerative functional} if
there exists $\{\tau_j^n\}$ such that
\begin{flushleft}
(i) 
$(\mathcal{K}^n_j,l_j^n)$ is independent to $\mathbb{F}^n_{\tau_j^n}$ for 
each $j=1,2,\dots$, \\
(ii) $(\mathcal{K}^n_j,l_j^n)$,
 $j=1,2,\dots$ are identically distributed.
\end{flushleft}
An additive functional of an ergodic diffusion is a
regenerative functional. See Fukasawa~\cite{F2008} for the details.
Let $K^n$ be a 
$d$-dimensional regenerative functional and put
$\mathcal{\bar{K}}^n_j = (\mathcal{K}^n_{j,l^n_j}, l_j^n)$
for $j=0,1,\dots$. 
Notice that
$\mathcal{\bar{K}}^n_j$, $j \geq 1$
is an iid sequence and independent of
$\mathcal{\bar{K}}^n_0$.
Assume that
$\mathrm{Var}^n[\mathcal{\bar{K}}^n_j]$
exists and is of rank $d^{\prime} + 1$ with $1\leq d^\prime \leq d$
for all $j \geq 1$. 
Without loss of generality, assume that there exists
a $d^{\prime}$-dimensional iid sequence $G_j^n$, $j \geq 1$ such that
the variance matrix of $(G_j^n,l_j^n)$ is of full rank and that
\begin{equation} \label{lgn}
\mathcal{\bar{K}}^n_j = (G_j^n, R_j^n,l_j^n)
\end{equation}
with a $d-d^\prime$ dimensional sequence $R^n_j$.
Put
\begin{equation*}
\begin{split}
&m_L^n = E^n[l_1^n], \ \ m_G^n =  E^n[G_1^n],
\ \  m_R^n =  E^n[R_1^n],
\\
& \mu^n =  (\mu^n_k) = (m_G^n,m_R^n)/m_L^n,
\end{split}
\end{equation*}
and 
\begin{equation*}
\mathbb{K}^n_j = (\mathbb{G}^n_j,l_j^n), \ \  
\mathbb{G}^n_j = G_j^n - l_j^n m_G^n/m_L^n, \ \ j \in \mathbb{N}.
\end{equation*}
Due to the definition, it is not difficult to see a
law of large numbers holds: 
\begin{equation*}
K^n_T /T \to \mu^n
\end{equation*}
in probability as $T \to \infty$. Further, a central limit theorem
\begin{equation*}
\sqrt{T}(K^n_T/T - \mu^n) \Rightarrow \mathcal{N}(0, V^n)
\end{equation*}
holds with a suitable matrix $V^n$.
Our aim here is to give a refinement of this central limit 
theorem. More precisely, for a given function 
$A^n:\mathbb{R}^d \to \mathbb{R}$ and a positive sequence 
$T_n \to \infty$, we  present a valid
approximation of the distribution of
\begin{equation*}
\sqrt{T_n}(A^n(K_{T_n}^n/T_n) - A^n(\mu^n))
\end{equation*}
up to $O(T_n^{-1})$ as $n \to \infty$.
As far as considering this form, we can assume
without loss of generality that 
$E^n[|R^n_j|] = 0$ 
for all $j \geq 1$ in (\ref{lgn}). 
Put 
\begin{equation*}
(\mu^n_{k,l}) = \mathrm{Var}^n[\mathbb{G}^n_1]/m^n_L, \ \ 
\rho^n = (\rho^n_k) = \mathrm{Cov}^n[\mathbb{G}^n_1, l_1^n]
\end{equation*}
and
\begin{equation*}
\mu_{k,l,m}^n = (\kappa_{k,l,m}^n - \rho_k^n\mu_{l,m}^n - 
\rho_l^n\mu_{m,k}^n - \rho_m^n\mu_{k,l}^n)/m_L^n,
\end{equation*}
where $(\kappa_{k,l,m}^n)$ is the third moments
of $\mathbb{G}_1^n$.
\begin{assu} \label{assl}
It holds that
\begin{equation*}
\inf_{n\in \mathbb{N}} m^n_L > 0.
\end{equation*}
\end{assu}
\begin{assu}  \label{assm}
For $\xi = (d^{\prime} + 2) \vee 4$, it holds that
\begin{equation*}
\sup_{n \in \mathbb{N}} \left\{
E^n[|\mathcal{\bar{K}}^n_0|^2] 
+ E^n[|\mathbb{K}^n_1|^\xi]
+ E^n\left[
\int_{\tau^n_1}^{\tau^n_2} |\mathcal{K}_{1,t}^n|^2 \mathrm{d}t
\right] \right\} < \infty.
\end{equation*}
\end{assu} 
Under Conditions~\ref{assl} and \ref{assm}, the sequences
$\mu^n$, $(\mu^n_{k,l})$, $(\mu^n_{k,l,m})$ are bounded 
in $n \in \mathbb{N}$.
\begin{assu} \label{asss}
Let $\Psi^n$ be the characteristic function of $\mathbb{K}_1^n$:
\begin{equation*}
\Psi^n(u) = E^n[\exp\{iu\cdot \mathbb{K}^n_1\}].
\end{equation*}
It holds
\begin{equation*}
\sup_{|u| \geq b, n \in \mathbb{N} }|\Psi^n(u)| < 1
\end{equation*}
for all $b > 0$ and
there exists $\eta \geq 1$ such that
\begin{equation*}
\sup_{n \in \mathbb{N}}
\int_{\mathbb{R}^{d^{\prime}+1}} |\Psi^n(u)|^\eta \mathrm{d}u < \infty.
\end{equation*}
\end{assu}
Note that under Conditions~\ref{assm} and \ref{asss}, it holds
\begin{equation*}
0 < \inf_{|a| = 1, n \in \mathbb{N}} 
E^n[|a \cdot \mathbb{K}^n_1|^2]
\leq
\sup_{|a| = 1, n \in \mathbb{N}} 
E^n[|a \cdot \mathbb{K}^n_1|^2] < \infty,
\end{equation*}
that is, the largest and smallest eigenvalues of the variance matrix
of $\mathbb{K}_1^n$ is bounded and bounded away from $0$ in 
$n \in \mathbb{N}$.

Let $B_n(\zeta)= \{x \in  \mathbb{R}^d; |x - \mu^n|<\zeta \}$ for
$\zeta > 0$, 
\begin{equation*}
a^n_i =\partial_i A^n(\mu^n), 
\ \ a^n_{i,j} = \partial_i \partial_j A^n(\mu^n),
\ \  1\leq i,j  \leq d
\end{equation*}
for a given function $A^n:\mathbb{R}^d \to \mathbb{R}$ which is twice
differentiable at the point $\mu^n$ and
\begin{equation*}
a^n = (a_k^n) \in \mathbb{R}^d,\ \ 
v^n  =\sum_{k,l=1}^{d^{\prime}} \mu^n_{k,l} a^n_k
a^n_l.
\end{equation*}

\begin{assu} \label{assh}
There exists $\zeta>0$ such that 
\begin{enumerate}
\item 
$A^n :\mathbb{R}^d \to \mathbb{R}$ is
four times continuously differentiable on $B_n(\zeta)$ 
for all $n$, 
\item 
all the derivatives up to fourth order 
are bounded on $B_n(\zeta)$ uniformly in $n$,
\item  it holds that
\begin{equation*}
0 < \inf_{n\in\mathbb{N}} v^n \leq
\sup_{n\in \mathbb{N}}v^n< \infty.
\end{equation*}
\end{enumerate}
\end{assu}
Denote by $\iota$ the natural inclusion:
$\mathbb{R}^{d^{\prime}} \ni v \mapsto (v,0,\dots,0) \in \mathbb{R}^d$.
\begin{thm}
\label{EE}
Let $M$ be a positive constant and $\mathcal{B}_M$ be the set of
Borel functions on $\mathbb{R}$ 
which are bounded by $M$.
Under Conditions \ref{assl}, \ref{assm}, \ref{asss} and \ref{assh},
it holds that
\begin{equation*}
\sup_{H \in \mathcal{B}_M, n \in \mathbb{N}}
T_n \left|E^n[H(\sqrt{T_n}(A^n(K^n_{T_n}/T_n) - A^n(\mu^n)))]
- \int H(z)q^n(z)dz\right| < \infty,
\end{equation*}
where $q^n$ is defined as
\begin{equation} \label{qn}
q^n(z) = \phi(z;v^n)
+ T_n^{-1/2}\left\{A^n_1 q_1(z;v^n) + \frac{A^n_3}{6} q_3(z;v^n)\right\}
\end{equation}
with $\phi(z;v^n)$ being the normal density with mean $0$ and variance $v^n$,
\begin{equation*}
 q_1(z;v^n) = - \partial\phi(z;v^n), \ \ 
 q_3(z;v^n) = - \partial^3\phi(z;v^n),
\end{equation*}
and 
\begin{equation} \label{eecoef}
\begin{split}
&A^n_1 =  \frac{1}{2}\sum_{k,l=1}^{n^{\prime}}
a^n_{k,l}\mu^n_{k,l}
 + a^n \cdot \left\{
E^n[K^n_{\tau^n_1}] + \frac{1}{m^n_L}
E^n\left[
\int_{\tau^n_1}^{\tau^n_2}K^n_t\mathrm{d}t
\right] - \frac{\iota(\rho^n)}{m^n_L}
\right\}, \\
&A^n_3 = \sum_{k,l,m=1}^{n^{\prime}} a^n_ka^n_la^n_m \mu^n_{k,l,m} + 
3\sum_{j,k,l,m=1}^{n^{\prime}}
a^n_ja^n_ka^n_{l,m}\mu^n_{j,l}\mu^n_{k,m}.
\end{split}
\end{equation}
\end{thm}
{\it Proof: }
The proof is a repetition of the proof of  Theorem~4.1 of
Fukasawa~\cite{F2008} with the aid of Theorem~\ref{BR} and so is omitted.
\hfill////

\section{Proof of Theorem~\ref{main}}

Here we give the proof of Theorem~\ref{main}.
We are considering (\ref{SV}) with 
$\varphi$, $\rho$, $U$ satisfying the condition stated
at the beginning of Section~3.
The initial value $(Z_0,X_0) =(0,0)$ and
the time to maturity $T$  are fixed. 
Now, to obtain a contradiction, let us suppose that the supremum in 
Theorem~\ref{main} is infinite. Then there exists a sequence 
$\theta_n \in \mathcal{C}(\gamma,\delta)$ such that
\begin{equation} \label{diverge}
\epsilon_n^{-2}\left|
\mathbb{E}[f(Z^n_T)]- \mathbb{E}[(1+p_n(N))
f(-\log(D)-\Sigma_n/2  + \sqrt{\Sigma_n}N)]
\right| \to \infty
\end{equation}
as $n\to \infty$, where $\epsilon_n = \epsilon_{\theta_n}$, 
$p_n = p_{\theta_n}$, $\Sigma_n = \Sigma_{\theta_n}$.
Put $\hat{b}_n = \epsilon_{n}^2 b_{\theta_n}$,
$\hat{c}_n = \epsilon_n c_{\theta_n}$  and denote by 
$\mathbb{E}^n_x$ the expectation operator with respect to 
the law of $\hat{X}^n$ determined by
the stochastic differential equation
\begin{equation*}
\mathrm{d}\hat{X}^n_t = 
\hat{b}_n(\hat{X}^n_t) \mathrm{d}t + 
 \hat{c}_n(\hat{X}^n_t) \mathrm{d}\hat{W}^1_t, 
\ \ \hat{X}^n_0 = x \in \mathbb{R},
\end{equation*}
where $\hat{W}^1$ is a standard Brownian motion.
It is easy to see that the law of $X = \{X_t\}$ in $(\ref{SV})$
is the same as that of 
$\left\{\hat{X}^n_{t/\epsilon_n^2} \right\}$ with $\hat{X}^n_0= 0$.
Hence, the law of $Z^n = Z^{\theta_n}$ is the same as that of 
\begin{equation*}
 -\log(D)-\frac{1}{2}\Sigma_n + 
 \sqrt{\Sigma_n} \sqrt{T_n}A^n(K^n_{T_n}/T_n)
\end{equation*}
under $\mathbb{E}^n_0$, where
\begin{equation} \label{setA}
T_n = \frac{T}{\epsilon_n^2}, \ \ 
A^n(x,y) = \frac{\sqrt{T}y - T_n^{-1/2}x/2}{\sqrt{\Sigma_n}}, \ \ 
h_n = T \varphi^2 -\Sigma_n,
\end{equation}
\begin{equation*}
K^n_t = \left(
\int_0^t h_n(\hat{X}^n_s)\mathrm{d}s,
\int_0^t \varphi(\hat{X}^n_s) \left[
\rho(\hat{X}^n_s) \mathrm{d}\hat{W}^1_s + \sqrt{1-\rho(\hat{X}^n_s)^2}
 \mathrm{d}\hat{W}^2_s
\right] \right)
\end{equation*}
and $(\hat{W}^1,\hat{W}^2)$ is a 2-dimensional standard Brownian
motion.
By the strong Markov property, $K^n$ is a regenerative functional
in the sense given in the previous section with the stopping times
$\{\tau^n_j\}$  defined as
\begin{equation} \label{seq}
\tau^n_0 = 0, \ \ 
\tau^n_{j+1} = \inf\left\{t > \tau^n_j; \hat{X}^n_t = x^n_0,
\sup_{s\in [\tau^n_j,t]}\hat{X}^n_s \geq  x^n_1 \right\},
\end{equation}
with an arbitrarily fixed point $(x^n_0,x^n_1) \in \mathbb{R}^2$ with
$x^n_0 <  x^n_1$.
Let us take $x^n_0=x$, $x^n_1 = x+a$ with $(x,a)$ which 
satisfies Condition~\ref{assint}; recall that 
$\theta_n \in \mathcal{C}(\gamma,\delta)$, so that we can find such
a pair $(x,a)$ for each $n$.

Put $s_n = s_{\theta_n}$, $\pi_n = \pi_{\theta_n}$ and 
$\Pi_n = \Pi_{\theta_n}$.  
To verify all the conditions for Theorem~\ref{EE} to hold, 
we use the following more-or-less known identities.
The first one is that
\begin{equation}\label{ergodistr}
\Pi_n[g] = \frac{1}{\mathbb{E}_{x^n_0}[\tau^n_1]}\mathbb{E}\left[
\int_0^{\tau^n_1}g(\hat{X}_t^n)\mathrm{d}t
\right]
\end{equation}
for all integrable function $g$; see e.g., Skorokhod~\cite{Sko}. Section~3.1. 
The second one is Kac's moment formula~\cite{FS}: 
for given a positive Borel function $g$, define 
\begin{equation*}
G_g^{k}(y;z) = \mathbb{E}^n_y\left[
\int_0^{\tau(z)}g(\hat{X}^n_t)G^{k-1}_g(\hat{X}^n_t;z)\mathrm{d}t 
\right]
\end{equation*}
recursively for $k \in \mathbb{N}$, 
where $y,z \in \mathbb{R}$, $G^0_g(y;z) \equiv 1$ and
\begin{equation*}
\tau(z) = \inf\{ t>0;\hat{X}^n_t = z\}.
\end{equation*} 
Then, it holds that
\begin{equation}\label{Kac}
\mathbb{E}^n_y\left[ \left|
\int_0^{\tau(z)}|f(\hat{X}^n_t)|\mathrm{d}t \right|^k 
\right] = k! G^k_{|f|}(y;z)
\end{equation}
for any $y,z \in \mathbb{R}$. 
The third one is that
\begin{equation*}
G^1_{g}(y;z) =
 2 \int_{y}^{z}(s_n(z)-s_n(x))g(x)\Pi_n(\mathrm{d}x) + 
2(s_n(z)-s_n(y))\int_{-\infty}^{y}g(x)\Pi_n(\mathrm{d}x)
\end{equation*}
if $y \leq z$, and
\begin{equation*}
G^1_{g}(y;z)=  2(s_n(y)-s_n(z))\int_{y}^{\infty}g(x)\Pi_n(\mathrm{d}x)
 + 2 \int_{z}^{y}(s_n(x)-s_n(z))g(x)\Pi_n(\mathrm{d}x)
\end{equation*}
if $y > z$. See Skorokhod~\cite{Sko}, Section~3.1 for the details.

\begin{lem}  \label{lem51}

Condition~\ref{assl} holds.
\end{lem}
{\it Proof: } By the strong Markov property and
the above identities,
\begin{equation*}
m^n_L = \mathbb{E}[\tau^n_2-\tau^n_1]= 
\mathbb{E}^n_{x^n_0}[\tau(x^n_1)] +  
\mathbb{E}^n_{x^n_1}[\tau(x^n_0)] =  
2(s_n(x^n_1)-s_1(x^n_0)).
\end{equation*}
The result then follows from Condition~\ref{assint}. \hfill////

\begin{lem}\label{lem52}
Condition~\ref{assm} holds.
\end{lem}
{\it Proof: }
By the Burkholder-Davis-Gundy inequality and
the strong Markov property, it suffices to show
\begin{equation*}
\sup_{n \in \mathbb{N}}\mathbb{E}^n_{y}\left[
|\tau(z)|^4 + 
\left|
\int_0^{\tau(z)}|h_n(\hat{X}^n_t)|\mathrm{d}t
\right|^4 + \left|\int_0^{\tau(z)} 
\varphi(\hat{X}^n_t)^2 \mathrm{d}t \right|^2
\right] < \infty
\end{equation*}
for $(y,z)=(0,x^n_0)$, $(y,z) = (x^n_0,x^n_1)$ and 
$(y,z) = (x^n_1,x^n_0)$.
We only need to show
\begin{equation} \label{opphi}
\sup_{n \in \mathbb{N}}\mathbb{E}^n_{y}\left[
\left|\int_0^{\tau(z)} (1+
\varphi(\hat{X}^n_t)^2) \mathrm{d}t \right|^4
\right] < \infty.
\end{equation}
because $h_n = T\varphi^2 -\Sigma_n$,
\begin{equation*}
\Sigma_n = T\Pi_n[\varphi^2] = \frac{T}{m^n_L}
\mathbb{E}^n_{x^n_0}\left[\int_0^{\tau^n_1} 
\varphi(\hat{X}^n_t)^2\mathrm{d}t\right]
\end{equation*}
and $\inf_nm^n_L > 0$ by Lemma~\ref{lem51}.
By Condition~\ref{assgam}, we have
\begin{equation*}
s^{\prime}_n(w)g_k(v)
\pi_n(v) \leq \frac{1}{\delta}e^{(k+1)\gamma_\pm|v|-
(4\gamma_\pm + \delta)|v-w|)}
\end{equation*}
for $g_k(v)=(1+\varphi(v)^2)\exp(k\gamma_\pm|v|)$, $k\in\mathbb{Z}$
if $|v|\geq|w|$ and $vw\geq0$,
where $\gamma_{\pm} = \gamma_+$ if $v \geq 0$ and
$\gamma_{\pm}=\gamma_-$ otherwise. 
Hence, by Condition~\ref{assint},
 there exists a constant $C$ (independent of $n$) 
such that 
\begin{equation*}
G^1_{g_k}(u;z) \leq C e^{(k+1)\gamma_{\pm}|u|}.
\end{equation*}
for any $u \in \mathbb{R}$ as long as $k \leq 3$, 
where $z = x^n_0$ or $z=x^n_1$.
This inequality implies (\ref{opphi}) with the aid of (\ref{Kac}).
\hfill////

\begin{lem} \label{lem53}
Condition~\ref{asss} holds.
\end{lem}
{\it Proof: }  The proof is lengthy so is deferred to Section~6. \hfill////

\begin{lem}\label{lem54}
Condition~\ref{assh} holds.
\end{lem}
{\it Proof: } Note that
\begin{equation*}
0 < \inf_{n \in \mathbb{N}} \Sigma_n \leq \sup_{n\in\mathbb{N}}
\Sigma_n < \infty
\end{equation*}
by Lemmas~\ref{lem51}, \ref{lem52} and \ref{lem53}.
The first two properties are then obvious from (\ref{setA}).
To see the third, notice that
\begin{equation} \label{vn}
\begin{split}
v^n =& \frac{T}{m^n_L \Sigma_n} \mathbb{E}
\left[\int_{\tau^n_1}^{\tau^n_2} \varphi^2(\hat{X}^n_t) \mathrm{d}t
\right] \\
&+\frac{\epsilon_n}{m^n_L \Sigma_n}
\mathbb{E}
\left[\int_{\tau^n_1}^{\tau^n_2} 
\psi_n(\hat{X}^n_t)\varphi(\hat{X}^n_t)\rho(X^n_t) \mathrm{d}t
\right] +O(\epsilon_n^2)\\
=& 1+\frac{\epsilon_n}{\Sigma_n} 
\Pi_n[\psi_n\varphi\rho] + O(\epsilon_n^2)
\end{split}
\end{equation}
and $\Pi_n[\psi_n\varphi\rho] = O(1)$,
in the light  of Lemma~\ref{lem52}, where
\begin{equation}\label{psin}
\psi_n(y) = 2s^\prime_n(y)\hat{c}_n(y)\int_{-\infty}^yh_n(w)
\Pi_n(\mathrm{d}w).
\end{equation}
Here we used the fact that
\begin{equation*}
0 = \int_{\hat{X}^n_{\tau^n_1}}^{\hat{X}^n_{\tau^n_2}}
\frac{\psi_n(x)}{\hat{c}_n(x)}\mathrm{d}x =
\int_{\tau^n_1}^{\tau^n_2}\psi_n(\hat{X}^n_t)\mathrm{d}\hat{W}^1_t 
+ \int_{\tau^n_1}^{\tau^n_2}h_n(\hat{X}^n_t)\mathrm{d}t,
\end{equation*}
which follows from the It$\hat{\text{o}}$-Tanaka formula.
\hfill////

Now we are ready to apply Theorem~\ref{EE}.
In the light of Lemma~3 of Fukasawa~\cite{F2008} and Lemma~\ref{lem52}, 
we have $A^n_1 = O(\epsilon_n)$.
Further, by the It$\hat{\text{o}}$-Tanaka formula and Lemma~\ref{lem52}, we obtain
\begin{equation*}
\begin{split}
&\mathbb{E}^n_{x^n_0}\left[
\left\{ \int_0^{\tau^n_1} \varphi(\hat{X}^n_t)
\left\{
\rho(\hat{X}^n_t)\mathrm{d}\hat{W}^1_t + 
\sqrt{1-\rho(\hat{X}^n_t)}\mathrm{d}\hat{W}^2_t
\right\}\right\}^3 \right]
\\
&= 3
\mathbb{E}^n_{x^n_0}\left[
\int_0^{\tau^n_1} \varphi(\hat{X}^n_t)
\left\{
\rho(\hat{X}^n_t)\mathrm{d}\hat{W}^1_t + 
\sqrt{1-\rho(\hat{X}^n_t)}\mathrm{d}\hat{W}^2_t
\right\} \int_0^{\tau^n_1} \varphi(\hat{X}^n_t)^2\mathrm{d}t \right]
\\
&= 3
\mathbb{E}^n_{x^n_0}\left[
\int_0^{\tau^n_1} \varphi(\hat{X}^n_t)
\rho(\hat{X}^n_t)\mathrm{d}\hat{W}^1_t \int_0^{\tau^n_1}
 \varphi(\hat{X}^n_t)^2\mathrm{d}t \right]
\\
&= 3\frac{\mathbb{E}^n_{x^n_0}[\tau^n_1]}{T}\left\{-
\Pi_n[\varphi \rho \psi_n] +
\frac{\Sigma_n}{\mathbb{E}^n_{x^n_0}[\tau^n_1]}
\mathbb{E}^n_{x^n_0} \left[ \tau^n_1
\int_0^{\tau^n_1}\varphi(\hat{X}^n_t)\rho(\hat{X}^n_t)\mathrm{d}\hat{W}^1_t
\right]
\right\},
\end{split}
\end{equation*}
where $\psi_n$ is defined by (\ref{psin}).
This implies
\begin{equation*}
A^n_2 = -\frac{3\Pi_n[\varphi \rho \psi_n]}{\Pi_n[\varphi^2]^{3/2}T} 
+ O(\epsilon_n) = 
\frac{6\alpha_n \sqrt{T_n}}{\sqrt{\Sigma_n}} + O(\epsilon_n),
\end{equation*}
where $\alpha = \alpha_n$ is defined as
(\ref{poly}) with $c = c_{\theta_n} = \hat{c}_n/\epsilon_n$,
$\Sigma = \Sigma_n$ and $\Pi = \Pi_n$.
Since
\begin{equation*}
\begin{split}
\phi(z;v^n) =& \phi(z;1) + \frac{\epsilon_n}{\Sigma_n} 
\Pi_n[\psi_n\varphi\rho]  \frac{\partial}{\partial v}\phi(z;v)
\Big|_{v = 1} + O(\epsilon_n^2)\\
=& \phi(z;1) - \alpha_n
\phi(z;1)(z^2-1) + O(\epsilon_n^2)
\end{split}
\end{equation*}
by (\ref{vn}), we conclude $q^n(z) = \phi(z;1)(1+p_n(z))+ O(\epsilon_n^2)$.
Hence we obtain a contradiction to (\ref{diverge}).

\section{Proof of Lemma~\ref{lem52}}
Here we prove that the characteristic function $\Psi^n(u)$ of
\begin{equation*}
\left(
\tau^n_1, \int_0^{\tau^n_1}h_n(\hat{X}^n_t)\mathrm{d}t,
\int_0^{\tau^n_1}\varphi(\hat{X}^n_t)\left[
\rho(\hat{X}^n_t)\mathrm{d}\hat{W}^1_t + \sqrt{1-\rho(\hat{X}^n_t)^2}
\mathrm{d}\hat{W}^2_t\right]
\right)
\end{equation*}
under $\mathbb{E}^n_{x^n_0}$ satisfies the inequalities
of Condition~\ref{asss}.
By the strong Markov property,
it suffices to prove the same inequalities for 
the characteristic function $\hat{\Psi}^n(u)$ of
\begin{equation*}
\left(
\tau(x^n_1), \int_0^{\tau(x^n_1)}h_n(\hat{X}^n_t)\mathrm{d}t,
\int_0^{\tau(x^n_1)}\varphi(\hat{X}^n_t)\left[
\rho(\hat{X}^n_t)\mathrm{d}\hat{W}^1_t + \sqrt{1-\rho(\hat{X}^n_t)^2}
\mathrm{d}\hat{W}^2_t\right]
\right)
\end{equation*}
under $\mathbb{E}^n_{x^n_0}$ instead of $\Psi^n(u)$.

Note that $Y^n := s_n(\hat{X}^n)$ 
is a local martingale by the It$\hat{\text{o}}$-Tanaka 
formula, so that there exists a standard Brownian motion $B^n$ such that
$Y^n = B^n_{\langle Y^n\rangle}$
by the martingale representation theorem. 
Under $\mathbb{E}^n_{x^n_0}$, $B^n_0 = s_n(x^n_0)$.
Note also that
\begin{equation*}
\mathrm{d}Y^n_t = 
s_n^\prime(\hat{X}^n_t) \hat{c}_n(\hat{X}^n_t) \mathrm{d}\hat{W}^1_t
= \sqrt{\frac{s_n^\prime}{\pi_n}}(\hat{X}^n_t) 
\mathrm{d}\hat{W}^1_t = \frac{1}{\sigma_n(Y^n_t)} \mathrm{d}\hat{W}^1_t,
\end{equation*}
where $\sigma_n(y) = \sqrt{\pi_n(s_n^{-1}(y))}
/\sqrt{s_n^\prime(s_n^{-1}(y))}$.
It follows that
\begin{equation*}
\begin{split}
&\int_0^{\tau} g(\hat{X}^n_t)\mathrm{d}t = 
\int_0^{\langle Y^n \rangle_{\tau}} 
g(s_n^{-1}(B^n_u))\sigma_n(B^n_u)^2 \mathrm{d}u, \\
&\int_0^{\tau} g(\hat{X}^n_t)\mathrm{d}\hat{W}^1 = 
\int_0^{\langle Y^n \rangle_{\tau}} 
g(s_n^{-1}(B^n_u))\sigma_n(B^n_u) \mathrm{d}B^n_u
\end{split}
\end{equation*}
for every finite stopping time $\tau$ and locally bounded Borel function
$g$.
When considering the hitting time $\tau = \tau(x^n_1)$ of $\hat{X}^n$, 
we have
\begin{equation}\label{tauhat}
\langle Y^n \rangle_{\tau} = \hat{\tau}_n :=
\inf\{ s > 0; B^n_s = s_n(x^n_1)\}.
\end{equation}
Put $y^n_i=s_n(x^n_i)$ for $i=0,1$ and 
$y^n_{-1} = s_n(2x^n_0-x^n_1)$.
Notice that by definition, 
\begin{equation*}
\inf_{n\in\mathbb{N}}|y^n_1-y^n_0| > 0,\ \ 
\inf_{n\in\mathbb{N}}|y^n_0-y^n_{-1}| > 0,\ \ 
\sup_{n\in\mathbb{N}}|y^n_1-y^n_{-1}| < \infty. 
\end{equation*}

\begin{lem}\label{Bor}
Let $B^n$ be a standard Brownian motion with $B^n_0 = y^n_0$
and define $\hat{\tau}_n$ as (\ref{tauhat}).
Let $\Lambda$ be a set and 
$g_n(\cdot,\lambda):\mathbb{R} \to \mathbb{R}$ 
be a sequence of Borel functions  for each $\lambda \in \Lambda$ with
\begin{equation}\label{supgn}
\sup_{\lambda \in \Lambda, n \in \mathbb{N},
v \in [y^n_{-1},y^n_1]} |g_n(v,\lambda)| < \infty.
\end{equation}
Then there exist positive constants $a_1$ and $a_2$ such that
for all  $\lambda \in \Lambda$ and $n \in \mathbb{N}$,
the distribution of
\begin{equation*}
\int_0^{\hat{\tau}_n} g_n(B^n_t,\lambda)\mathrm{d}t
\end{equation*}
is infinite divisible 
with L\'evy measure $L$
satisfying for all $z>0$,
\begin{equation*}
L((-\infty,-z]) \vee
 L((z,\infty)) \leq a_1 + 
\frac{a_2}{\sqrt{z}}.
\end{equation*}
Moreover,
\begin{itemize}
\item if  there exists a sequence of intervals
$I_n \subset [y^n_{0},y^n_1]$ such that
\begin{equation*}
\inf_{n\in \mathbb{N}} |I_n| > 0, \ \ 
\inf_{\lambda \in\Lambda, n \in \mathbb{N}, v \in I_n} 
g_n(v,\lambda) > 0,
\end{equation*}
then there exist positive constants $a_3$ and $a_4$ such that
\begin{equation*}
-a_3 + \frac{a_4}{\sqrt{z}} \leq L((z,\infty))
\end{equation*}
holds for all $z > 0$, $\lambda \in \Lambda$ and $n \in \mathbb{N}$,
\item 
if  there exists a sequence of intervals
$I_n \subset [y^n_{0},y^n_1]$ such that
\begin{equation*}
\inf_{n\in \mathbb{N}} |I_n| > 0, \ \ 
\sup_{n \in \mathbb{N}, \lambda \in \Lambda, v \in I_n} 
g_n(v,\lambda) <  0,
\end{equation*}
then there exist another positive constants $a_3$ and $a_4$ such that
\begin{equation*}
-a_3 + \frac{a_4}{\sqrt{|z|}} \leq L((\infty,z])
\end{equation*}
holds for all $z < 0$, $\lambda \in \Lambda$ and $n \in \mathbb{N}$.
\end{itemize}

\end{lem}
{\it Proof: }
This can be proved by the same argument as in the proof of
Lemma~3 of  Borisov~\cite{Bor}.
\hfill////

\begin{lem} \label{Ind}
Let $(B^n,\check{B}^n)$ be a 2-dimensional standard Brownian motion 
with $B^n_0 = y^n_0$
and define $\hat{\tau}_n$ as (\ref{tauhat}).
Let $g_n$ be a sequence of locally bounded Borel functions with
\begin{equation*}
\sup_{n \in \mathbb{N}, v \in [y^n_{-1},y^n_1]}|g_n(v)| < \infty, \ \ 
\inf_{n \in \mathbb{N}, v \in [y^n_{-1},y^n_1]}|g_n(v)| > 0.
\end{equation*}
Then there exist positive constants $a_1, a_2, a_3, a_4$ such that
the distribution of
\begin{equation*}
\int_0^{\hat{\tau}_n} g_n(B^n_t)\mathrm{d}\check{B}^n_t
\end{equation*}
is infinite divisible 
with L\'evy measure $L$ satisfying
\begin{equation*}
-a_1 + \frac{a_2}{z}  \leq L((-\infty,-z]) = 
 L((z,\infty)) \leq a_3 + 
\frac{a_4}{z}
\end{equation*}
for all $z>0$, $n \in \mathbb{N}$.
\end{lem}
{\it Proof: }
Put $\Delta_n = y^n_1-y^n_0$.
Let $\tau_{i/m}$, $i=1,\dots,m$ 
be the times at which $B^n$
first attains the levels $y^n_0 + \Delta_ni/m$ respectively.
Put
\begin{equation*}
J_n = \sum_{i=1}^mJ^{mi}_n, \ \ 
J^{mi}_n=
\int_{\tau_{(i-1)/m}}^{\tau_{i/m}}g_{n}(B^n_t)\mathrm{d}\check{B}^n_t.
\end{equation*}
Note that $J^{mi}_n$,
$i=1,\dots,m$ are independent
by the strong Markov property.  Besides,
$\{J^{mi}_n\}_{1\leq i\leq m}$ 
is a null array for each $n \in \mathbb{N}$ since
for all $\epsilon > 0$,
\begin{equation*}
\sup_{1\leq i \leq m} P[
|J^{mi}_n| > \epsilon ] \leq 
P[M \tau_{1/m} N^2 > \epsilon^2] + A_m
\end{equation*}
which converges to $0$ as $m \to \infty$, where
$M$ is a constant, $N$ is a standard normal
 variable independent of $B^n$, and
\begin{equation}\label{Am}
A_m = \sup_{1\leq i \leq m, n \in \mathbb{N}}
P\left[\left\{\inf_{\tau_{(i-1)/m}\leq t \leq \tau_{i/m}} 
B^n_t \leq y^n_{-1} \right\}  \right] = 
\frac{1}{m}\frac{1}{\inf_{n \in \mathbb{N}}|y^n_{0}-y^n_{-1}|}.
\end{equation}
Hence, $J_n$ is infinite divisible for each $n \in \mathbb{N}$.
Denoting by $L$ its L\'evy measure, it holds that
for every continuity point $z > 0$,
\begin{equation}\label{levymeasp}
\lim_{m \to \infty} \sum_{i=1}^m P[J^{mi}_n > z] 
= L((z,\infty))
\end{equation}
and for every continuity point $z < 0$ of $L$, 
\begin{equation}\label{levymeasm}
\lim_{m \to \infty} \sum_{i=1}^m P[J^{mi}_n \leq  z] 
= L((-\infty,z]),
\end{equation}
for which see e.g., Feller~\cite{Feller}, XVII.7.
Observe that for $z > 0$, 
\begin{equation*}
\begin{split}
&P\left[J^{mi}_n  < -z\right] = P\left[J^{mi}_n  > z\right] \\
&\leq A_m + \int_0^{\infty}
\int_{z/\sqrt{M t}}^{\infty}\phi(y;1)\mathrm{d}y \frac{\Delta_n}
{m \sqrt{2\pi  t^3}} \exp\left\{
-\frac{\Delta_n^2}{2tm^2}\right\}\mathrm{d}t
\\ 
&= A_m + 
\int_0^{\infty}
\int_{z/\sqrt{M}}^{\infty}
\frac{\Delta_n}
{ 2\pi m t^2} \exp\left\{
-\frac{\Delta_n^2 + m^2u^2}{2tm^2}\right\}\mathrm{d}u \mathrm{d}t
\\
&= A_m + 
\int_{z/\sqrt{M}}^{\infty}
\int_0^{\infty} \frac{\Delta_n}
{ 2\pi m } \exp\left\{
-\frac{s}{2}\left\{u^2 + \frac{\Delta_n^2}{m^2} \right\}\right\}
\mathrm{d}s \mathrm{d}u \\
&=A_m + \frac{1}{\pi}\int_{mz/(\Delta_n\sqrt{M})}^{\infty}
\frac{\mathrm{d}v}{1+v^2}
\end{split}
\end{equation*}
where $M$ is a constant. Hence, by L'Hopital's rule and (\ref{levymeasp}),
\begin{equation*}
L((-\infty,-z]) = 
 L((z,\infty)) 
\leq a + \frac{\sup_{n\in \mathbb{N}}\Delta_n\sqrt{M}}{\pi z}
\end{equation*}
with a constant $a > 0$.
By the same calculation, we have also
\begin{equation*}
L((-\infty,-z]) = 
 L((z,\infty)) 
\geq- a + \frac{\inf_{n \in \mathbb{N}}\Delta_n\sqrt{M^\prime}}{\pi z}
\end{equation*}
with another constant $M^\prime > 0$.
\hfill////

\begin{lem} \label{whole}
Let $(B^n,\check{B}^n)$ be a 2-dimensional standard Brownian motion 
with $B^n_0 = y^n_0$
and define $\hat{\tau}_n$ as (\ref{tauhat}).
Let $\Lambda$ be a set, 
$g_{n,1}(\cdot,\lambda)$ be Borel functions 
for each $\lambda \in \Lambda$,
$g_{n,2}$ be Borel functions which are absolutely continuous 
on $[y^n_{-1},y^n_1]$ respectively, and 
$g_{n,3}$ be Borel functions   with
\begin{equation*}
\sup_{\lambda \in \Lambda, n \in \mathbb{N},  v \in [y^n_{-1},y^n_1]}
|g_{n,1}(v,\lambda)|  \vee |g_{n,2}(v)|\vee |g^\prime_{n,2}(v)| \vee |g_{n,3}(v)| 
\vee \frac{1}{|g_{n,3}(v)| }  < \infty.
\end{equation*}
Assume that there exists a sequence of intervals 
$I_n \subset [y^n_0,y^n_1]$ with 
\begin{equation*}
\inf_{n \in \mathbb{N}} |I_n| > 0
\end{equation*}
such that
\begin{equation}\label{or}
\inf_{\lambda \in \Lambda, n \in \mathbb{N}, v \in I_n}
g_{n,1}(v,\lambda) > 0 \ \ \text{ or }
\sup_{\lambda \in \Lambda, n \in \mathbb{N},v \in I_n}
g_{n,1}(v,\lambda)<  0 
\end{equation}
holds.
Denote by $\hat{g}_n(\cdot;u,\lambda)$  the characteristic function  of
$J$ defined as
\begin{equation*}
J = u_1\int_0^{\hat{\tau}_n} g_{n,1}(B^n_t,\lambda)\mathrm{d}t +
u_2\int_0^{\hat{\tau}_n} g_{n,2}(B^n_t)\mathrm{d}B^n_t + 
u_2\int_0^{\hat{\tau}_n} g_{n,3}(B^n_t)\mathrm{d}\check{B}^n_t,
\end{equation*}
where $u = (u_1,u_2) \in \mathbb{R}^2$ with $|u| = 1$.
Then, there exists a constant $C \in (0,\infty)$ such that
for every $t \in \mathbb{R}$, it holds 
\begin{equation*}
\sup_{\lambda \in \Lambda, n \in \mathbb{N}, 
u;|u| = 1}|\hat{g}_n(t;u,\lambda)| \leq C e^{-\sqrt{|t|}/C}.
\end{equation*}
\end{lem}
{\it Proof:  }
Put $\Delta_n = y^n_1-y^n_0$ and
let $\tau_{i/m}$, $i=1,\dots,m$ 
be the times at which $B^n$
first attains the levels $y^n_0 + \Delta_ni/m$ respectively
as in the previous proof.
Put
\begin{equation*}
\begin{split}
&J^{mi,1}_n=  u_1\int_{\tau_{(i-1)/m}}^{\tau_{i/m}} 
g_{n,1}(B^n_t,\lambda)\mathrm{d}t + 
u_2 \int_{\tau_{(i-1)/m}}^{\tau_{i/m}}g_{n,2}(B^n_t)\mathrm{d}B^n_t,
\\ 
& J^{mi,2}_n=
u_2
 \int_{\tau_{(i-1)/m}}^{\tau_{i/m}}g_{n,3}(B^n_t)\mathrm{d}\check{B}^n_t
\end{split}
\end{equation*}
and $J^{mi}_n = J^{mi,1}_n + J^{mi,2}_n$.
By the same argument as before, we conclude that
$J$ is infinitely divisible for each $n\in \mathbb{N}$, 
$\lambda \in \Lambda_n$ and $u \in \mathbb{R}^2$.
We have (\ref{levymeasp}) and (\ref{levymeasm})
with its L\'evy measure $L$.
Notice that
\begin{equation}\label{gn2}
\int_{\tau_{(i-1)/m}}^{\tau_{i/m}}g_{n,2}(B^n_t)\mathrm{d}B^n_t = 
\int_{y^n_0 + (i-1)\Delta_n/m}^{y^n_0 + i\Delta_n/m}
g_{n,2}(y)\mathrm{d}y - \frac{1}{2}
\int_{\tau_{(i-1)/m}}^{\tau_{i/m}} g_{n,2}^\prime(B^n_t)\mathrm{d}t
\end{equation}
on the set
\begin{equation*}
\left\{\inf_{\tau_{(i-1)/m}\leq t \leq \tau_{i/m}} 
B^n_t > y^n_{-1} \right\} 
\end{equation*}
by the It$\hat{\text{o}}$-Tanaka formula.
Since, for example,
\begin{equation*}
\begin{split}
&P[J_n^{mi} > z] \geq
P\left[J_n^{mi,2} > 2z \right] -
P[J_n^{mi,1}\leq -z],\\
&P[J_n^{mi} > z] \leq
P\left[J_n^{mi,2} > z/2 \right] +
P[J_n^{mi,1} > z/2],
\end{split}
\end{equation*}
there exist positive constants
$a_i$, $i=1,2, \dots,6$ such that
\begin{equation}\label{zp1}
-a_1 - \frac{a_2}{\sqrt{z}} + 
 \frac{a_3|u_2|}{z} 
\leq L((z,\infty)) \leq a_4 + 
\frac{a_5}{\sqrt{z}}
+ \frac{a_6|u_2|}{z}
\end{equation}
for all $z>0$ and 
\begin{equation*}
-a_1 - \frac{a_2}{\sqrt{|z|}} + 
 \frac{a_3|u_2|}{|z|} 
\leq L((-\infty,z]) \leq a_4 + 
\frac{a_5}{\sqrt{|z|}}
+ \frac{a_6|u_2|}{|z|}
\end{equation*}
for all $z < 0$ by Lemmas~\ref{Bor}, \ref{Ind} and 
(\ref{Am}), (\ref{levymeasp}), (\ref{levymeasm}).

In case we have the first inequality in (\ref{or}), if
\begin{equation*}
|u_2| \leq \beta_0 :=
\frac{1}{\sqrt{2}} \wedge \frac{\beta_1}{2\beta_2}
\end{equation*}
with 
$\beta_1 = \inf_{\lambda \in \Lambda, n \in \mathbb{N}, 
v \in I_n}g_{n,1}(v,\lambda)$ and
$\beta_2 = \sup_{n \in \mathbb{N}, v \in I_n}|g_{n,2}^\prime(v)|$,
then 
\begin{equation*}
\begin{split}
&\inf_{n \in \mathbb{N}, \lambda \in \Lambda_n, v \in I_n} 
u_1 g_{n,1}(v;\lambda) - u_2g^\prime_{n,2}(v)/2
\\&\geq \sqrt{1-|u_2|^2}\beta_1 - |u_2|\beta_2/2 \geq \beta_1/4 > 0,
\end{split}
\end{equation*}
so that by Lemma~\ref{Bor} and (\ref{gn2}),
\begin{equation*}
\lim_{m \to \infty} \sum_{i=1}^m P[J_n^{mi,1} > z] 
\geq -\tilde{a}_1 + \frac{\tilde{a}_2 }{\sqrt{z}}
\end{equation*}
for all $z>0$, 
where $\tilde{a}_i, i=1,2$ are positive constants.
In addition, we have
\begin{equation*}
P[J_n^{mi} > z] \geq P[J_n^{mi} > z; J_n^{mi,1} > z] \geq 
\frac{1}{2}P[J_n^{mi,1} > z]
\end{equation*}
for all $z > 0$.
Hence, when $|u_2| \leq \beta_0$,
there exist another constants $a^\prime_i$, $i=1,2$
such that 
\begin{equation}\label{zp2}
-a^{\prime}_1 +  \frac{a^{\prime}_2}{\sqrt{z}} 
\leq L((z,\infty)) \leq a_4 + 
\frac{a_5}{\sqrt{z}}
+ \frac{a_6|u_2|}{z}
\end{equation}
for all $z > 0$. 
Now, note that by the L\'evy-Khinchin expression, 
there exists a constant $\sigma^2 \geq 0$ such that
\begin{equation*}
\mathrm{Re} \log(\hat{g}_n(t;u,\lambda)) = -\sigma^2t^2/2 -2\int_{\mathbb{R}}
\sin^2(zt/2)L(\mathrm{d}z).
\end{equation*}
Take $z_1 > 0$ such that 
$z \mapsto  \sin^2(z)$ is increasing on 
$[0,z_1/2]$. Then, observe that for sufficiently small 
$z_0 \in (0,z_1)$, it holds that
\begin{equation*}
\frac{a_2^{\prime}/\sqrt{z_0} - a_5/\sqrt{z_1}}{a_6/z_1}
> \frac{a_2/\sqrt{z_0} + a_5/\sqrt{z_1}}{a_3/z_0 - a_6/z_1} > 0.
\end{equation*}
Fix such a point $z_0$ and take $\beta_3$ such that
\begin{equation*}
\frac{a_2^{\prime}/\sqrt{z_0} - a_5/\sqrt{z_1}}{a_6/z_1}
> \beta_3 >  \frac{a_2/\sqrt{z_0} + a_5/\sqrt{z_1}}{a_3/z_0 - a_6/z_1}.
\end{equation*}
Then we have for the case that $|u_2|\sqrt{t} \geq \beta_3$,
\begin{equation} \label{l1}
\begin{split}
&\int_{\mathbb{R}}\sin^2(zt/2)L(\mathrm{d}z) \\
& \geq
\int_{(z_0/|t|,z_1/|t|]} \sin^2(zt/2)L(\mathrm{d}z) \\
& \geq
\sin^2(z_0/2)L((z_0/|t|,z_1/|t|]) \\
& \geq \sin^2(z_0/2)\left\{
\left(\frac{a_3}{z_0} - \frac{a_6}{z_1}\right)|u_2||t| 
-\left(\frac{a_2}{\sqrt{z_0}} + \frac{a_5}{\sqrt{z_1}}\right)\sqrt{|t|}
- a_1-a_4 \right\}\\
&\geq
\sqrt{|t|}/C - \log(C)
\end{split}
\end{equation}
for sufficiently large constant $C$ by (\ref{zp1}).
Further, by (\ref{zp2}), we have
for the case that $|u_2|\sqrt{t} \leq \beta_3$,
\begin{equation*} \label{l2}
\begin{split}
&\sin^2(z_0/2)L((z_0/|t|,z_1/|t|]) \\
& \geq  \sin^2(z_0/2)\left\{
\left(\frac{a_2^{\prime}}{\sqrt{z_0}} - \frac{a_5}{\sqrt{z_1}}\right)
\sqrt{|t|} 
-\frac{a_6}{z_1}|u_2||t|
- a^{\prime}_1-a_4 \right\}\\
&\geq
\sqrt{|t|}/C - \log(C).
\end{split}
\end{equation*}

The same conclusion is obtained also in the case that we have 
the second inequality instead of the first in (\ref{or}).
For example, we define $\beta_1$ alternatively as
$\beta_1 = -\sup_{\lambda \in \Lambda, n \in \mathbb{N}, 
v \in I_n}g_{n,1}(v,\lambda)$ and observe
\begin{equation*}
 \lim_{m \to \infty} \sum_{i=1}^m P[J_n^{mi,1} \leq z] 
\geq -\tilde{a}_1 + \frac{\tilde{a}_2 }{\sqrt{|z|}}
\end{equation*}
for all $z<0$ with positive constants $\tilde{a}_1$, $\tilde{a}_2$
when $|u_2|\leq \beta_0$. Then use
\begin{equation*}
P[J_n^{mi} \leq z] \geq 
P[J_n^{mi} \leq z ; J_n^{mi,1} \leq z ] \geq
\frac{1}{2}P[J_n^{mi,1} \leq z]
\end{equation*}
to obtain
\begin{equation*}
-a^{\prime}_1 +  \frac{a^{\prime}_2}{\sqrt{|z|}} 
\leq L((\infty,z]) \leq a_4 + 
\frac{a_5}{\sqrt{|z|}}
+ \frac{a_6|u_2|}{|z|}
\end{equation*} 
for all $z < 0$. The rest is a straightforward translation.
\hfill////

Now we are ready to prove Lemma~\ref{lem52}.
By Petrov's lemma (see Petrov~\cite{Petrov}, p.10),
it suffices to prove that there exists
a constant $C \in (0,\infty)$ such that 
\begin{equation*}
|\hat{\Psi}^n(u)|\leq Ce^{-|u|/C},
\end{equation*}
where $\hat{\Psi}^n$ is 
what was defined at the beginning of this section.
Put
\begin{equation*}
\begin{split}
&g_{n,1}(v,\lambda) = (\lambda_1 + \lambda_2h_n(s_n^{-1}(v)))\sigma_n(v),
\\
&g_{n,2}(v) = \varphi(s_n^{-1}(v))\rho(s_n^{-1}(v))\sigma_n(v), \\
&g_{n,3}(v) = \varphi(s_n^{-1}(v))\sqrt{1-\rho(s_n^{-1}(v))^2}\sigma_n(v)
\end{split}
\end{equation*}
and
\begin{equation*}
M(\lambda) =
\sup_{n\in\mathbb{N},
v\in [y^n_{-1},y^n_1]} |g_{n,1}(v,\lambda))| \vee |g_{n,2}(v)| 
\vee |g_{n,2}^\prime(v)|
\vee |g_{n,3}(v)| \vee \frac{1}{|g_{n,3}(v)|}
\end{equation*}
for $\lambda = (\lambda_1,\lambda_2) \in \mathbb{S}$,
the 1-dimensional unit sphere.
It is not difficult to see that
for all $\lambda \in \mathbb{S}$, we have $M(\lambda) < \infty$
and that there exists a sequence of  intervals 
$I_n(\lambda) \subset [y^n_0,y^n_1]$ with 
\begin{equation*}
\inf_{n \in \mathbb{N}} |I_n(\lambda)| > 0
\end{equation*}
such that
\begin{equation*}
\inf_{n \in \mathbb{N}, v \in I_n(\lambda)}
g_{n,1}(v,\lambda) > 0 \ \ \text{ or }
\sup_{n \in \mathbb{N},v \in I_n(\lambda)}
g_{n,1}(v,\lambda)<  0 
\end{equation*}
holds. If the first inequality holds for $\lambda = \lambda_0$,
put
\begin{equation*}
m(\lambda_0) = \inf_{n \in \mathbb{N}, v \in I_n(\lambda_0)}
g_{n,1}(v,\lambda_0)
\end{equation*}
and
\begin{equation*}
\Lambda(\lambda_0) = 
\left\{
\lambda \in \mathbb{S}; 
\inf_{n \in \mathbb{N}, v \in I_n(\lambda_0)}
g_{n,1}(v,\lambda) > m(\lambda_0)/2, M(\lambda) < 2M(\lambda_0)
\right\}.
\end{equation*}
If the second inequality holds for $\lambda = \lambda_0$,
put
\begin{equation*}
m(\lambda_0) = \sup_{n \in \mathbb{N}, v \in I_n(\lambda_0)}
g_{n,1}(v,\lambda_0)
\end{equation*}
and
\begin{equation*}
\Lambda(\lambda_0) = 
\left\{
\lambda \in \mathbb{S} ;
\sup_{n \in \mathbb{N}, v \in I_n(\lambda_0)}
g_{n,1}(v,\lambda) < m(\lambda_0)/2, M(\lambda) < 2M(\lambda_0)
\right\}.
\end{equation*}
Now, notice that $\Lambda(\lambda_0), \lambda_0 \in \mathbb{S}$
is an open covering of $\mathbb{S}$, so that it has a finite subcovering
$\Lambda(\lambda_1), \cdots, \Lambda(\lambda_J)$.
For each $\lambda_j$, we can apply Lemma~\ref{whole} with
$\Lambda = \Lambda(\lambda_j)$ to obtain that there exists $C_j > 0$
such that
\begin{equation*}
\sup_{\lambda \in \Lambda(\lambda_j), u \in \mathbb{S},
n\in \mathbb{N}}|\hat{\Psi}^n(tu_1\lambda_1,tu_1\lambda_2,tu_2)|
\leq C_je^{-|t|/C_j}
\end{equation*}
for all $t \in \mathbb{R}$. Since $J < \infty$, we conclude that
there exists $C>0$ such that
\begin{equation*}
\sup_{\lambda \in \mathbb{S}, u \in \mathbb{S},
n\in \mathbb{N}}|\hat{\Psi}^n(tu_1\lambda_1,tu_1\lambda_2,tu_2)|
\leq Ce^{-|t|/C}
\end{equation*}
for all $t \in \mathbb{R}$, which completes the proof of Lemma~\ref{lem52}.

\end{document}